\documentclass[twocolumn]{autart}    
\usepackage{graphicx}          
\usepackage{amsmath,amssymb}
\usepackage{nccmath}

\begin{document}

\begin{frontmatter}
\runtitle{Optimal control over Markovian channels}  

\title{Optimal control over Markovian wireless communication channels under generalized packet dropout compensation\thanksref{footnoteinfo}} 

\thanks[footnoteinfo]{This paper was not presented at any IFAC meeting. Corresponding author: Y.~Zacchia Lun.}

\author[DISIM]{Yuriy Zacchia Lun}\ead{yuriy.zacchialun@univaq.it}, 
\author[DICEA]{Francesco Smarra}\ead{francesco.smarra@univaq.it}, and 
\author[DISIM]{Alessandro D'Innocenzo}\ead{alessandro.dinnocenzo@univaq.it}  

\address[DISIM]{Department of Information Engineering, Computer Science and Mathematics, University of L'Aquila, 67100 L'Aquila, Italy} 

\address[DICEA]{Department of Civil, Construction-Architectural \& Environmental Engineering, University of L'Aquila, 67100 L'Aquila, Italy} 
          

\begin{abstract}  
Control loops closed over wireless links greatly benefit from accurate estimates of the communication channel condition. To this end, the finite-state Markov channel model allows for reliable channel state estimation. This paper develops a Markov jump linear system representation for wireless networked control with persistent channel state observation, stochastic message losses, and generalized packet dropout compensation. With this model, we solve the finite- and infinite-horizon linear quadratic regulation problems and introduce an easy-to-test stability condition for any given infinite-horizon control law. We also thoroughly analyze the impact of a scalar general dropout compensation factor on the stability and closed-loop performance of a rotary inverted pendulum controlled remotely through a wireless link. Finally, we validate the results numerically via extensive Monte Carlo simulations, showing the benefits of the proposed control strategy.
\end{abstract}
\end{frontmatter}
\section{Introduction}
Wireless networked control systems receive considerable attention from industry and academia thanks to their mission-critical applications in industrial automation, intelligent transportation, telesurgery, and smart grids. See, e.g., \cite{park2018comm,eisen2019iot,pezzutto2020adaptive,liu2021iot} as an overview of significant recent advances in the wireless networked control system research. One of the central topics in this research area is estimation and control over fading channels, explored, e.g., in 
\cite{schenato2007proc,gupta2009data,heemels2010networked,goncalves2010signal,ding2011automatica,pajic2011tac,minero2013tac,quevedo2014tcst,fu2015optimal,zacchialun2019cdc,liu2022remote,impicciatore2022cdc,impicciatore2024tac}.

The performance of systems having their control loops closed over wireless links is strongly affected by the communication channels' stochastic behavior since wireless links are subject to path loss, shadowing, and fading when mobility is involved: this gives rise to time-varying message dropouts, message delays, and jitter \cite{goldsmith2005wireless}. Consequently, an accurate estimate of the channel condition for correctly modeling the stochastic properties of a wireless networked control system can significantly increase the control performance. 

The finite-state Markov channel (FSMC; see, e.g., \cite{sadeghi2008finite}) is a simple yet powerful analytical model that captures the dynamics of the wireless link. It is widely used for analyzing and designing telecommunications systems.
Despite the availability of the FSMC model, when dealing with the application level, packet dropouts dynamics are often modeled as realizations of a Bernoulli process \cite{schenato2007proc,hu2021smc}, which may result in an oversimplification of the complex communication subsystem dynamics and, thus, in an incorrect analysis of the control subsystem properties, for instance in terms of stability \cite{zacchialun2019cdc}.

\subsection{Study motivation and technical challenges}
Recently, \cite{impicciatore2024tac} introduced FSMC models into a wireless networked control framework for optimal output-feedback control, proving the validity of the separation principle, which allows for designing the optimal remote system state estimator and controller separately. It showed that due to the one-time-step delayed channel state observation by the remote controller, the optimal control over FSMC links problem represents the main challenge in wireless output-feedback control since for the same FSMC link characteristics, the remote system may be detectable but not controllable. One significant limitation of \cite{impicciatore2024tac} addressed in this paper is considering only the zero-input packet dropout compensation at the actuator end, which results in sub-optimal performance.
To fill this gap, this paper exploits an FSMC wireless link abstraction to model packet dropouts, as in \cite{zacchialun2020infocom}. 
Similarly to \cite{impicciatore2024tac} and \cite{10531715}, it considers a persistent channel state observation, providing a controller with the outcome of each control packet transmission and observed state channel state in a positive or negative acknowledgment message. However, this paper introduces a generalized control packet dropout compensator to the closed-loop system architecture. Specifically, the actuators apply an appropriately scaled last available control input when the communication link corrupts a transmitted control message. This approach encompasses both zero-input and hold-input packet dropout compensations as notable cases, and it is particularly suitable for less performant actuators incapable of immediately zeroing the control inputs. Markedly, this article differs from previous works on generalized control packet dropout compensation, such as \cite{moayedi2013networked}, \cite{fu2015optimal}, and \cite{lu2018switching}, in considering the FSMC to govern the packet dropout dynamics instead of the Bernoulli process, resulting in a much more general and complex problem setup presented in Section \ref{sec:model}. 
The generalized control packet dropout compensation strategy is straightforward, intuitive, and structurally simple, making it appealing from the implementation and cost perspectives \cite{fu2015optimal}. Other compensators can, in principle, be constructed based on a static or dynamic, linear or nonlinear combination of several past controls at different previous time instants. Still, such compensators would be more costly, complex, and energy-consuming.

Considering the generalized control packet dropout compensation within the wireless networked control framework under the FSMC link model presents several major technical challenges and offers a significant increase in control performance, as demonstrated in Sections \ref{subsec:dropout-factor} and \ref{subsec:comparison}. The first challenge lies within the controller specification and the resulting closed-loop system model. It requires answering whether the control gain should depend only on the last observed channel state or whether some additional information, such as the current packet dropout interval, is necessary. This paper shows in Section \ref{subsec:lqr-model} that the channel state information is sufficient for selecting the optimal control gain at run time, thus limiting the complexity and the related processing delay. However, the closed-loop system model must consider the FSMC-state-dependent packet dropout interval process, requiring different novel perspectives for control design and stability analysis, as detailed in Sections \ref{subsec:lqr-model} and \ref{subsec:stability-model}. An additional challenge of the finite-horizon optimal controller design also comes from the stochastic nature of the packet dropout interval, which requires the controller to balance the time horizon and the possible number of consecutive control message dropouts at each time step, as detailed in Section \ref{sec:optimal-control}. For the infinite-horizon case, establishing the ergodicity of the closed-loop system's discrete dynamics is another significant technical challenge addressed in this paper in Proposition \ref{prop:ergodicity}. None of these challenges were tackled in the current wireless networked control system literature. Only successfully addressing them allows us to cast and solve optimally the discrete-time linear stochastic systems with information between sensors, controllers, and actuators carried by FSMCs in a Markovian jump linear system (MJLS) framework and to rely on its fundamental principles.

\subsection{Related works}
The MJLS theory has been extensively used to investigate feedback control problems over lossy links.

Several works used a simplified Gilbert channel modeled by a Markov chain with two states. 
Specifically, \cite{seiler2005h} derived necessary and sufficient linear matrix inequality (LMI) conditions for synthesizing the optimal $H_{\infty}$ controller, assuming the controller is collocated with the actuator so that only the sensor measurements may be lost. 
\cite{kawka2006stability} analyzed the stability and performance of a channel-state-independent (CSI) controller with $N$-step packet loss compensation under one step of actuation delay, where transmitting packets containing multiple control inputs increases actuation delay and packet error probability due to larger packet size. 
\cite{xie2009stability} investigated stability properties of sampled-data networked linear systems under CSI control.
\cite{you2010minimum} derived the minimum data rate for mean square (MS) stabilizability under an arbitrary CSI quantized state-feedback (SF) control policy. 
\cite{minero2013tac} extended this result to a time-varying data rate modeled by a finite-state Markov chain.
\cite{mo2013lqg} solved finite- and infinite-horizon CSI linear-quadratic-Gaussian (LQG) control problems and proved the validity of the separation principle. 
\cite{battilotti2019lq} extended the finite-horizon CSI result to a non-Gaussian setting, assuming that the moments of the noise sequences up to the fourth order are known.
\cite{okano2017stabilization} derived conditions for MS stability of uncertain autoregressive systems whose state and input parameters vary within given intervals and characterized limitations on data rate, packet loss probabilities, and magnitudes of parametric uncertainty under CSI control.
Finally, \cite{li2021opt-zero} and \cite{li2021opt-hold} solved the finite- and infinite-horizon linear quadratic regulation (LQR) problems for discrete-time systems with a known constant input delay for the zero-input and hold-input control packet dropout compensation, respectively.
However, the two-state Markov chain model used in all the works above cannot represent a nontrivial FSMC \cite{impicciatore2024tac}. 
Moreover, disregarding the Markov channel state in the control design leads to simpler but more conservative solutions that apply only to a subset of plants stabilizable with channel-state-dependent 
controllers \cite{impicciatore2024tac}.

Some works used a finite-state Markov chain to describe the evolution of a packet error burst length without considering the observed communication channel state.
In particular, \cite{wu2007design} derived sufficient LMI conditions and corresponding control laws for the stochastic stability of noiseless systems with hold-input packet dropout compensation at the controller and actuator end. 
\cite{wang2013h} presented an $H_{\infty}$ SF control method and sufficient conditions for controllers ensuring stochastic stability with a specific disturbance attenuation level.
Other works relied on a finite-state Markov chain description of all possible packet arrival sequences of a certain length. These models also did not consider the Markov channel state information. 
A notable example is \cite{peters2019predictive}, which addressed the predictive control design problem for networked systems subject to packet loss in the controller-to-actuator link.
Additional noteworthy works characterized the packet losses in combination with other significant concerns, such as jamming attacks on communication links \cite{cetinkaya2017networked} and random delays \cite{xu2022channel}. Still, the Markov channel state information was neglected.

\subsection{Contribution statement}
The main contributions of this paper are the following.
\begin{enumerate}
    \item We present a practical Markov jump linear system model of discrete-time linear stochastic systems with a generalized control message dropout compensation over lossy actuation links modeled by FSMCs (see Section \ref{sec:mjls}).
    \item We solve finite- and infinite-horizon LQR problems, providing the SF gains that depend on the FSMC state observed with one time-step delay (Sections \ref{sec:optimal-control} and \ref{sec:inf-hor-sol} and Remark \ref{rem:future}).
    \item We introduce a necessary and sufficient stability condition for any given infinite-horizon SF control law that requires computing the spectral radius of a stability verification matrix and, thus, is easy to check (see Section \ref{sec:stability}).
    \item We validate all theoretical results numerically via extensive Monte Carlo simulations of a rotary inverted pendulum controlled remotely over a wireless link, where the link model relies on an accurate representation of a realistic wireless communication protocol. Furthermore, we provide a comparative analysis of proposed and existing LQR strategies and illustrate the impact of a scalar general dropout compensation factor on stability and closed-loop performance (see Section \ref{sec:examples}).
\end{enumerate}

\section{Model and problem formulation}\label{sec:model}
Consider a discrete-time linear stochastic system with intermittent control messages due to lossy communication and generalized dropout compensation:
\begin{align}\label{eq:state}
\begin{cases}
x_{k+1} = A x_{k} + B u_{k}^{} + w_{k}, \\
u_{k}^{} = \delta_{k} u_{k}^{c} + (1-\delta_{k}) \mathit{\Phi} u_{k-1}^{}.
\end{cases}
\end{align}
We use a standard notation, where $x_k\in\mathbb{R}^{n_x}$ and $u_k^{}\in\mathbb{R}^{n_u}$ are a system state and a control input to actuators, while $A$ and $B$ are state and input matrices of appropriate size, respectively; $w_k\in\mathbb{R}^{n_x}$ is a white Gaussian process noise having zero mean and covariance matrix $\Sigma_w$. 
\begin{assum}\label{assum:1}
    The process noise $w_k$ is independent of the initial state $x_0$ and the binary stochastic variable $\delta_k$.
\end{assum}
In the expression of $u_k$, $\delta_k$ models the packet loss between the controller and the actuators, and  $u_k^c\!\in\!\mathbb{R}^{n_u}$ is the desired control input computed by the remote controller. If the control message is correctly delivered, {$u_k^{}=u_k^c$}; otherwise, if lost, the actuators apply the last available control input multiplied by a matrix $\mathit{\Phi}=\bigoplus_{i=1}^{n_u} \phi_i$, where $\oplus$ indicates the direct sum that produces a diagonal matrix with the elements $\phi_i \in [0,1]$ on the main diagonal. This generalized approach covers both the zero-input (with all $\phi_i=0$) and hold-input (with all $\phi_i=1$) control message dropout compensations as its particular cases. 

\begin{figure}
\begin{center}
\includegraphics[width=\columnwidth]{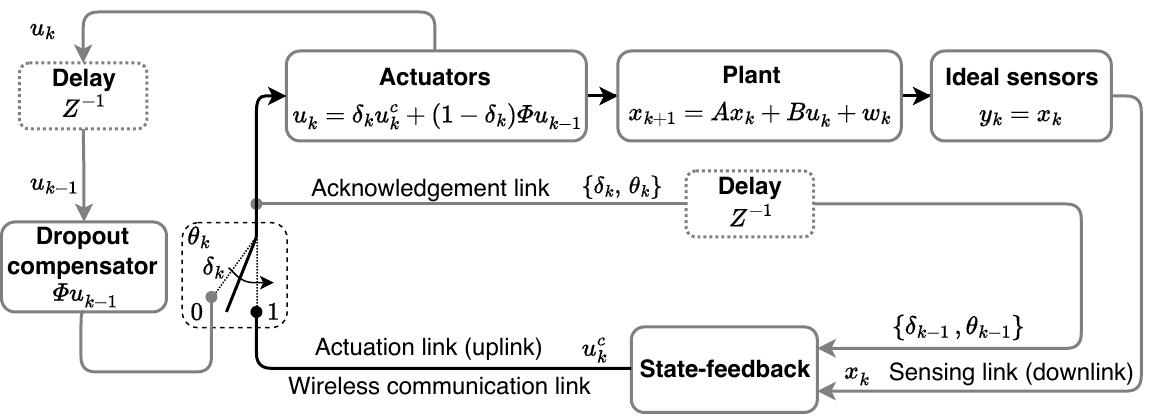}
\caption{Closed-loop system architecture. A wireless link delivers SF control inputs to actuators. The receiver measures the link state $\theta_k$ and communicates it with the transmission outcome $\delta_k$ to the controller. The actuators apply the generalized packet dropout compensation.}\label{fig:architecture}
\end{center}
\end{figure}

\begin{figure}
\begin{center}
\includegraphics[width=\columnwidth]{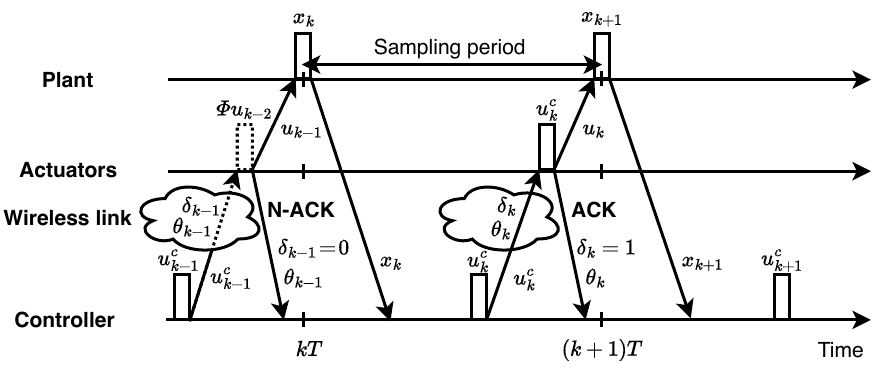}
\caption{A timing diagram for a closed-loop system with a wireless actuation link. In this example, the transmission corrupts the control packet containing $u^{c}_{k-1}$, as indicated by a dotted line. The receiver detects an error, discards the message, and sends the N-ACK containing the channel state $\theta_{k-1}$. Actuators compensate by applying a scaled version of the previous input signal, i.e. $u_{k-1}=\mathit{\Phi} u_{k-2}$ depicted by a dotted rectangular. The solid line in a cloud indicates successful wireless transmission of the control message $u^{c}_{k}$ to be applied by actuators so that $u_{k}=u_{k}^{c}$. Thus, the receiver sends the ACK also containing the channel state, i.e., $\theta_{k}$.}\label{fig:timing}
\end{center}
\end{figure}

We capture the dynamics of the control packet loss process $\left\{\delta_{k}\right\}$ in wireless links via an FSMC \cite{sadeghi2008finite}
abstraction\footnote{See, e.g., \cite{zacchialun2020infocom} for a procedure producing a consistent and accurate FSMC model suitable for a wireless industrial automation scenario.}. 
The wireless channel state is the output of a discrete-time Markov chain (DTMC) taking values in a finite set $\mathcal{S} \triangleq \{ s_i \}_{i=1}^N$. This state conditions the probability of successful packet delivery and packet loss:
\begin{align}\label{eq:p-delta}
\begin{aligned}
\mathbb{P}(\delta_k = 1 \mid \theta_{k} = s_i) & = \hat{\delta}_{i}, \\
\mathbb{P}(\delta_k = 0 \mid \theta_{k} = s_i) & = 1 - \hat{\delta}_{i}.
\end{aligned}   
\end{align}
In other words, we relate each state $s_i$ of the FSMC to a binary symmetric channel 
with error probability $1-\hat{\delta}_{i}$. The probabilities of transitions between FSMC states 
\begin{equation}\label{eq:p-ij}
p_{ij} \triangleq \mathbb{P}(\theta_{k} = s_j \mid \theta_{k-1} = s_i) \geq 0, \quad 
	\sum\nolimits_{j=1}^N p_{ij} = 1.
\end{equation}
We gather them into the channel transition probability matrix (TPM)
\begin{equation}\label{eq:Pc}
P_{c}^{} \triangleq \left[p_{ij}\right]_{i,j=1}^{N}.
\end{equation}
\begin{assum}\label{assum:2}
    The DTMC $\{\theta_k\}$ is ergodic. 
\end{assum}
\begin{assum}\label{assum:3}
    The process noise $\{w_k\}$ and the DTMC $\{\theta_k\}$ are independent.
\end{assum}
In many practical wireless communication scenarios, for instance, when networks rely on IEEE 802.15.4-compatible hardware\footnote{The typical networking protocols for wireless industrial automation, such as WirelessHART, ISA100.11a, and Zigbee PRO 2015, use IEEE 802.15.4.}, the FSMC state is available to receivers \cite{zacchialun2020infocom}.
So, as in \cite{zacchialun2019cdc}, we assume that a controller may observe Markov channel states via positive and negative acknowledgments (ACKs and N-ACKs), which become available only after the current decision on the control gain to apply has been made and sent through the wireless link since the actual transmission outcome is unknown in advance. 
Formally, the information set available to the controller is
\begin{equation}\label{eq:info-set}
\mathcal{I}_{k} = \left\{
	\left(x_{t}\right)_{t=0}^{k}, 
	\left(\delta_{t-1}\right)_{t=1}^{k}, 
	\left(\theta_{t-1}\right)_{t=1}^{k} \right\}.
\end{equation}
It relies on an idealistic assumption of accessing all system state variables over an error free link made only to streamline the presentation. We have already shown in \cite{impicciatore2024tac} how to design the optimal output-feedback controller by solving the optimal SF control and optimal filtering problems separately for a networked control scenario with the zero-input compensation strategy for the packet losses in the wireless links conveying sensing and actuation data. Since the acknowledgment messages transporting the related transmission outcome and channel state information are short and unlikely to be corrupted, we also make an idealistic assumption that all the ACKs are always successfully delivered\footnote{One can readily relax it 
by introducing an additional binary stochastic variable to model the successful delivery of the ACK that would evolve according to an FSMC describing the corresponding wireless link.}.
Fig.~\ref{fig:architecture} shows the closed-loop system architecture, and Fig.~\ref{fig:timing} provides the related timing diagram. This paper solves the following design problem.

\emph{Problem statement:}
Design a finite- or infinite-horizon controller for the system \eqref{eq:state}
that fully exploits its 
information set \eqref{eq:info-set}. Thus, the SF gain should depend on the Markov channel state observed with one time-step delay\footnote{See Remark \ref{rem:1} for the technical motivation for the control gain dependence on the channel state but not the observed values of the packet loss process.
Intuitively, the system state $x_k$ contains the transmission outcome information encoded in $\left(\delta_{t-1}\right)_{t=1}^{k}$.}:
\begin{equation}\label{eq:control}
u_{k}^{c} 
= K_{(k,\theta_{k-1})}x_k.
\end{equation}
Specifically, consider a  time horizon $T\in\mathbb{Z}^{>} \cup \{\infty\}$ and denote by $\mathcal{U}_{T}$ a set of all control inputs satisfying \eqref{eq:control}.
In the following, $\mathbb{Z}^{>}$ and $\mathbb{Z}^{\geq}$ will indicate the sets of all positive and nonnegative integers, respectively.
Given a positive semi-definite state-weighting matrix $Q \in \mathbb{R}^{n_x}$ and a positive-definite input-weighting matrix $R\in \mathbb{R}^{n_u}$, the optimal linear quadratic regulator minimizes the following costs. Denote by $\mathbb{E}$ the expectation. For $T<\infty$,
\begin{align}\label{eq:cost-def-fin}
\begin{aligned}
    & J_{T}^{}(x_0,\mathcal{U}_T) = \\
    & ~~ \mathbb{E}\left(\sum\nolimits_{k=0}^{T-1}\left(x_k^{\top} Q x_k + u_k^{\top}Ru_k\right) + x_T^{\top} Q x_T \mid \mathcal{I}_0 \right),
\end{aligned}
\end{align}
where $^{\top}$ indicates the transpose.
Notice that the cost \eqref{eq:cost-def-fin} weights the inputs to actuators and, thus, accounts for both the dropout compensation gain $\mathit{\Phi}$ and the desired control inputs $u_k^c \in \mathcal{U}_T$, depending on the realizations of the stochastic process $\{\delta_k\}$. So, the following 
optimization problem defines the finite-horizon LQR paradigm:
\begin{equation}\label{eq:control-fin}
    \check{\mathbf{u}}_T^c  = \mathop{\mathrm{arg\,min}}_{\mathbf{u}_T^c \in \mathcal{U}_T} J_{T}^{}(x_0,\mathcal{U}_T),
\end{equation}
where $\mathbf{u}_T^c \triangleq \left(u_t^c\right)_{t=0}^{T-1}$ is a sequence of control inputs.
For $T=\infty$, consider the long-run average cost
\begin{equation}\label{eq:cost-def-inf}
    J_{\infty}^{}(x_0,\mathcal{U}_{\infty}) = \limsup_{T\to\infty} \frac{1}{T} J_{T}^{}(x_0,\mathcal{U}_T).
\end{equation}
Then, the solution of the following optimization problem defines the infinite-horizon LQR paradigm:
\begin{equation}\label{eq:control-inf}
    \check{\mathbf{u}}_{\infty}^c  = \mathop{\mathrm{arg\,min}}_{\mathbf{u}_{\infty}^c \in \mathcal{U}_{\infty}} J_{\infty}^{}(x_0,\mathcal{U}_{\infty}).
\end{equation}
To analytically solve problems \eqref{eq:control-fin} and \eqref{eq:control-inf} in the FSMC setting, we first represent system \eqref{eq:state} in the Markovian jump linear system (MJLS) framework, as detailed next. 

\section{Markovian jump system model for control}\label{sec:mjls}
This section derives
an MJLS model that accounts for a generalized control message dropout compensation over a lossy actuation link modeled as an FSMC. After characterizing the system states in opportune time instances in Section \ref{subsec:time-instances}, it defines the system's discrete states and the related transition probabilities for the LQR in Section \ref{subsec:lqr-model}. 

\subsection{System states in packet delivery time instances}\label{subsec:time-instances}
The intuition behind the MJLS derivation is that system \eqref{eq:state} has a unique and convenient representation in the time instances in which actuators successfully receive messages from controllers. These time instances differ by a nonnegative number of consecutive control message dropouts governed by an FSMC.
Thus, we count the number of consecutive control message dropouts observable by a controller at a given time step $k$ in a stochastic variable $\mathit{\Delta}_{k}$. The value of $\mathit{\Delta}_{k}$ increments by one when the reception of a control message is not acknowledged and resets to zero otherwise. Formally,
\begin{equation}\label{eq:z-k}
\mathit{\Delta}_{k}=(1-\delta_{k-1})(\mathit{\Delta}_{k-1}+1).
\end{equation}
By iterating \eqref{eq:z-k} over multiple time steps, we have that 
\begin{equation}\label{eq:zl}
\mathit{\Delta}_{k}=\ell\Leftrightarrow \delta_{k-1-\ell}=1 \land 
	\delta_{k-t}=0 ~ \forall t\in\mathbb{Z}^{>} : t\leq \ell.
\end{equation}
Notice that if $\ell=0$, then $\left\{t\right\}_{t=1}^{0} = \emptyset$, meaning that \eqref{eq:zl} becomes $\mathit{\Delta}_{k}=0\Leftrightarrow \delta_{k-1}=1$. 

Let $\mathcal{T}$ be a set of time instances in which actuators successfully receive the controller's messages, i.e., 
\begin{subequations}\label{eq:tau} 
\begin{equation}\label{eq:calT}
    \mathcal{T} \triangleq \left\{ k : \delta_k = 1 \right\}_{k\in \mathbb{Z}^{\geq}} = \left\{ \tau_{(m)} \right\}_{m\in \mathbb{Z}^{\geq}}.
\end{equation}
From \eqref{eq:z-k}, \eqref{eq:zl}, and \eqref{eq:calT}, for all $m\in\mathbb{Z}^{\geq}$,
\begin{equation}\label{eq:deltatau}
    \tau_{(m)} \in \mathcal{T}  \Rightarrow \delta_{\tau_{(m)}} = 1 \Rightarrow \mathit{\Delta}_{\tau_{(m)}+1} = 0,
\end{equation}
\begin{equation}\label{eq:tau+}
    \tau_{(m+1)} = \tau_{(m)} + 1 + \mathit{\Delta}_{\tau_{(m+1)}}.
\end{equation}
\end{subequations}
For notational convenience, for any $k,n\in \mathbb{Z}^{\geq}$, let
\begin{equation}\label{eq:calPsi}
\mathit{\Psi}_{(n)} \triangleq \sum\nolimits_{j=0}^{n}A^{j}B\mathit{\Phi}^{n-j},
\end{equation}
\begin{equation}\label{eq:calGamma}
\mathit{\Gamma}_{(k,n)} \triangleq \sum\nolimits_{j=0}^{n}A^{n-j} w_{k+j}^{}.
\end{equation}
Then, the following proposition provides the foundation for the technical results of the paper.
\begin{prop}\label{prop:equiv}
    The system \eqref{eq:state} with control packet loss process $\left\{\delta_{k}\right\}$ governed by an FSMC described by \eqref{eq:p-delta}--\eqref{eq:Pc} and an arbitrary control strategy satisfying \eqref{eq:control} is trace-equivalent to the following system, where the components and time instances are defined by \eqref{eq:z-k}--\eqref{eq:calGamma}.
    \begin{equation}\label{eq:xtau}
    \begin{cases}
    \begin{aligned}
        \mathit{\Delta}_{\tau_{(m+1)}} &= n \in \mathbb{Z}^{\geq}  \Rightarrow  \forall h \in \mathbb{Z}^{\geq} : h \leq n, \\
    \end{aligned} \\
    \begin{aligned}
         x_{\tau_{(m)}+1+h}^{} &= A^{h+1} x_{\tau_{(m)}}^{} \!\!+\! \mathit{\Psi}_{(h)}u_{\tau_{(m)}}^{} \!\!+\! \mathit{\Gamma}_{(\tau_{(m)},h)},\!\!\\
    \end{aligned} \\
    \begin{aligned}
         u_{\tau_{(m)}+h}^{} &= \mathit{\Phi}^{h} K_{(\tau_{(m)},\theta_{\tau_{(m)}-1})} x_{\tau_{(m)}}^{}.
    \end{aligned}
    \end{cases}
    \end{equation}
    \begin{pf}
        By construction, \eqref{eq:z-k}--\eqref{eq:xtau} describe the dynamics of the system \eqref{eq:state}--\eqref{eq:Pc} constrained by \eqref{eq:control}.
    \end{pf}
\end{prop}
\begin{rem}
Proposition \ref{prop:equiv} does not address transition probabilities between $\tau_{(m)}$ and $\tau_{(m+1)}$, and the trace equivalence means that for any given initial system state $x_{0}$, control law satisfying \eqref{eq:control}, 
and realization of $\left(\delta_{t}\right)_{t=0}^{\tau_{m}+n}$ and $\left(w_{t}\right)_{t=0}^{\tau_{m}+n}$, the states $x_{\tau_{(m)}}$ and $x_{\tau_{(m)}+1+h}$ obtained from \eqref{eq:state} and \eqref{eq:xtau} will be the same $\forall h\in \mathbb{Z}^{\geq}$ such that $h\leq n$. Taken alone, \eqref{eq:xtau} presents all possible effects of packet error bursts and a control command satisfying \eqref{eq:control} under the generalized packet dropout compensation on the system state dynamics when considered for arbitrary values $n \in \mathbb{Z}^{\geq}$ of $\mathit{\Delta}_{\tau_{(m+1)}}$ in the controller's perspective. On the contrary, for any observed sequence of time instances in $\mathcal{T}$, \eqref{eq:xtau} reproduces the dynamics of \eqref{eq:state} in these time instances from the actuators' and plant's perspective. Notice from \eqref{eq:tau} that time instances in $\mathcal{T}$ must obey \eqref{eq:p-delta}--\eqref{eq:p-ij}.
\end{rem}
System \eqref{eq:xtau} formalizes that all the system states that precede a state with successful reception of a controller’s message depend on the previously received control command, the duration of a packet error burst following the last successful control message reception, the packet dropout compensation strategy, and the evolution of the process noise. 
Besides, \eqref{eq:info-set}, \eqref{eq:z-k}, and \eqref{eq:tau} make clear that the number of consecutive control message dropouts following a successful reception of a control command is unknown to a controller beforehand. However, the current number of consecutive control message dropouts $\mathit{\Delta}_{\tau_{(m)}}$ and the previous FSMC state $\theta_{\tau_{(m)}-1}$ are part of the controller's information set $\mathcal{I}_{\tau_{(m)}}$. To find the expression of the control gain, define the operational modes of a system \eqref{eq:xtau} from the controller's perspective as follows.

\subsection{System model for the LQR}\label{subsec:lqr-model}
Group together the current duration of a packet error burst and the last known wireless channel state in one augmented discrete state: $\eta_{k}^{} \triangleq (\mathit{\Delta}_{k},\theta_{k-1})$. From \eqref{eq:tau},
\begin{subequations}\label{eq:etatau}
\begin{equation}
    \eta_{\tau_{(m)}}^{} = (\mathit{\Delta}_{\tau_{(m)}},\theta_{\tau_{(m)}-1}),
\end{equation}
\begin{equation}
    \eta_{\tau_{(m+1)}}^{} = (\mathit{\Delta}_{\tau_{(m+1)}},\theta_{\tau_{(m)}+\mathit{\Delta}_{\tau_{(m+1)}}}),
\end{equation} 
\end{subequations}
where $\mathit{\Delta}_{\tau_{(m+1)}}$ indicates the time interval the transmitted control input may remain active. From the controller's perspective, $\eta_{\tau_{(m)}}^{}$ is known, while $\eta_{\tau_{(m+1)}}^{}$ is a random variable. In what follows, we derive the conditional probability of $\eta_{\tau_{(m+1)}}^{}$ with respect to $\eta_{\tau_{(m)}}^{}$.

Denote by $e_i$ the column vector of the standard basis of $\mathbb{R}^{N}$: all its components are zero except the $i$th, which equals one. 
Furthermore, store the probability of successful control packet delivery (or, conversely, of packet dropout) in a state of FSMC starting from a particular previous state in the matrix $P_{1}^{}$ (or, respectively, $P_{0}^{}$):
\begin{equation}\label{eq:P0and1}
    P_{1}^{} \triangleq \text{\small $[p_{ij}\hat{\delta}_{j}]_{i,j=1}^{N}$},~
    P_{0}^{} \triangleq \text{\small $[p_{ij}(1-\hat{\delta}_{j})]_{i,j=1}^{N}$} = P_{c}^{} - P_{1}^{}.
\end{equation}
Then, from \eqref{eq:p-delta}--\eqref{eq:Pc}, \eqref{eq:z-k}--\eqref{eq:tau}, \eqref{eq:etatau}, and \eqref{eq:P0and1}, the chain rule of the probability and independence of both $\delta_{k}$ and $\theta_{k}$ of $\delta_{k-t}$ for all $t \in \mathbb{Z}^{>}$,
\begin{align}\label{eq:zeta-long}
    & \mathbb{P}(\eta_{\tau_{(m+1)}}^{} = (n,s_j) \mid \eta_{\tau_{(m)}}^{} = (\ell,s_i)) = \\
    & \text{\small $\mathbb{P}(\delta_{\tau_{(m)}+t}=0 ~ \forall t\in\mathbb{Z}^{>} : t\leq n,\delta_{\tau_{(m)}+n+1}=1,\theta_{\tau_{(m)}+n} = s_j$} \notag \\
    & \text{\small $\mid \theta_{\tau_{(m)}-1} = s_i, \delta_{\tau_{(m)}}=1)$} = 
    \frac{e_i^{\top} P_{1}^{} P_{0}^{n} e_{j}^{}e_{j}^{\top} P_{1}^{} \mathbf{1}}{e_i^{\top} P_{1}^{}\mathbf{1}} \triangleq \zeta_{(i,n,j)}. \notag 
\end{align}
In \eqref{eq:zeta-long}, $\mathbf{1}$ indicates the column vector of appropriate size with all components equal to one.
\begin{rem}\label{rem:1}
The transition probabilities in \eqref{eq:zeta-long} are independent of $\mathit{\Delta}_{\tau_{(m)}}$.
Thus, all augmented-discrete-state-dependent control gains with the same last known FSMC state $\theta_{\tau_{(m)}-1}$ have identical probabilities of being received successfully and remaining active during a packet error burst. 
Hence, the optimal mode-dependent control gains should be the same for any given value of $\theta_{\tau_{(m)}-1}$ regardless of the current duration of a packet error burst\footnote{Otherwise, applying different control gains would produce the same cost, and the solution to the optimization problem would not be unique, which is not the case 
in our setting.}: designing an optimal mode-dependent controller would produce an optimal FSMC-state-dependent controller, and substituting $\theta_{\tau_{(m)}-1}$ with $\eta_{\tau_{(m)}}$ in \eqref{eq:xtau} would not alter the system's behavior under an optimal control law.
\end{rem}
For the notational convenience, let 
\begin{subequations}\label{eq:tp}
\begin{equation}\label{eq:varsigma}
    q_{i n} \triangleq \sum\nolimits_{j=1}^{N} \zeta_{(i,n,j)},\quad
    \varsigma_{(n,j)} \triangleq P_{0}^{n} e_{j}^{}e_{j}^{\top} P_{1}^{} \mathbf{1},
\end{equation}
i.e., the vector of probabilities of an $n$-length packet error burst that ends with the FSMC being in state $s_j$, and
\begin{equation}\label{eq:zeta}
    \zeta_{(i,n,j)} = \frac{e_i^{\top} P_{1}^{} \varsigma_{(n,j)}}{e_i^{\top} P_{1}^{}\mathbf{1}},
\end{equation}
\end{subequations}
a compact form of the transition probability \eqref{eq:zeta-long}.

Notice from \eqref{eq:p-delta}--\eqref{eq:Pc} and \eqref{eq:P0and1} that all the parameters in \eqref{eq:tp} are nonnegative and $\sum\nolimits_{n\in\mathbb{Z}^{\geq}}^{}\sum\nolimits_{j=1}^{N} \varsigma_{(n,j)} = \mathbf{1}$.
For big enough values $\hat{n}$ of a packet error burst $\mathit{\Delta}_{\tau_{(m+1)}}$, the probability $\varsigma_{(\hat{n},j)}$ becomes negligible and does not contribute to the summation above, i.e., there is a maximal number of consecutive control message dropouts $L \in \mathbb{Z}^{>}$ such that
$\sum\nolimits_{n=0}^{L}\sum\nolimits_{j=1}^{N} \varsigma_{(n,j)} \geq (1 - \epsilon) \mathbf{1}$,
where $\epsilon$ is arbitrarily small. 
In the numerical case studies using floating-point arithmetic, $\epsilon$ typically corresponds to the machine epsilon. Thus, the maximal
number of consecutive control message dropouts, $L$, results from the following optimization:
\begin{equation}\label{eq:L}
    L \triangleq \mathop{\mathrm{arg\,min}}_{\hat{n} \in \mathbb{Z}^{>}} \sum\nolimits_{n=0}^{\hat{n}}\sum\nolimits_{j=1}^{N} \varsigma_{(n,j)} \geq (1 - \epsilon) \mathbf{1},
\end{equation}
easily solvable by, e.g., the bisection method.
\begin{rem}\label{rem:future}
The system matrices from the MJLS model \eqref{eq:xtau}, \eqref{eq:etatau}, and \eqref{eq:zeta-long}--\eqref{eq:L} depend on the future, not present, operational mode $\eta_{\tau_{(m+1)}}$ forecast by the controller based on transition probabilities \eqref{eq:zeta}. Thus, the optimal mode-dependent SF control problem differs from the existing formulations presented, e.g., in \cite{costa2006discrete} and \cite{baras2008ifac} and provides a different, more complex solution averaging the effect of the control action through all possible packet error bursts.
\end{rem}
\begin{lem}\label{lemma:sys-lqr}
The system \eqref{eq:state}--\eqref{eq:p-ij} constrained by \eqref{eq:info-set} and \eqref{eq:control} is a MJLS described by \eqref{eq:z-k}--\eqref{eq:tp}.
\begin{pf}
    It follows from Proposition \ref{prop:equiv} and Remark \ref{rem:1}. 
\end{pf}
\end{lem}

\section{Optimal finite-horizon LQR}\label{sec:optimal-control}
This section provides the finite-horizon solution to the problem \eqref{eq:control-fin}, one of the main contributions of this paper.
In the finite-horizon setting, the control commands must balance the time horizon and possible number of consecutive control message dropouts at each time step. Specifically, on the one hand, a transmitted control command may remain active for a long time, up to the maximal number of consecutive control message dropouts $L$ defined by \eqref{eq:L}. Thus, the control command should average its effects over the $L+1$ discrete states of the system. On the other hand, near the time horizon end, the most recent control commands may never reach the actuators if the number of consecutive control message dropouts becomes greater than the remaining time horizon. Thus, at the end of the time horizon, the control commands average the effects over fewer states. The following theorem formalizes this concept.
\begin{thm}\label{thm:1}
The finite-horizon LQR solving the problem \eqref{eq:control-fin} for the system 
\eqref{eq:state}--\eqref{eq:p-ij} is given by 
\begin{equation}\label{eq:fh-u}
    \check{u}_{k}^{c} = K_{(k,\theta_{k-1})} x_k,
\end{equation}
where, $\forall k \in \mathbb{Z}^{\geq} : k < T$ and any value $s_i \in \mathcal{S}$ of $\theta_{k-1}$,
\begin{equation}\label{eq:fh-k}
    K_{(k,s_i)} = - \mathcal{B}_{(k,s_i)}^{-1}\mathcal{C}_{(k,s_i)},
\end{equation}
\vspace*{-6mm}
\begin{subequations}\label{eq:fh-x-a-b-c}
    \begin{align}\label{eq:fh-c}
        \mathcal{C}&_{(k,s_i)} = \sum\nolimits_{h=0}^{L-\xi_k}q_{ih}\sum\nolimits_{r=1}^{h} \mathit{\Psi}_{(r-1)}^{\top} Q A^{r} \,+ \\
        & \sum\nolimits_{h=0}^{L-\xi_k} 
        \sum\nolimits_{j=1}^{N} \zeta_{(i,h,j)} \mathit{\Psi}_{(h)}^{\top} \mathcal{X}_{(k+1+h,s_j)} A^{h+1} \notag ,   
    \end{align}
    \vspace*{-6mm}
    \begin{align}\label{eq:fh-b}
        \mathcal{B}&_{(k,s_i)} = R \,+ 
        \sum\nolimits_{h=0}^{L-\xi_k}q_{ih} \sum\nolimits_{r=1}^{h} \mathit{\Phi}^{r \top} R \mathit{\Phi}^r \,+ \\
        & \sum\nolimits_{h=0}^{L-\xi_k}q_{ih} \sum\nolimits_{r=1}^{h} \mathit{\Psi}_{(r-1)}^{\top} Q \mathit{\Psi}_{(r-1)}^{} \, + \notag \\
        & \sum\nolimits_{h=0}^{L-\xi_k} \sum\nolimits_{j=1}^{N} \zeta_{(i,h,j)} \mathit{\Psi}_{(h)}^{\top} \mathcal{X}_{(k+1+h,s_j)}  \mathit{\Psi}_{(h)}^{}, \notag
    \end{align}
    \begin{equation}\label{eq:fh-x}
        \mathcal{X}_{(k,s_i)} = \mathcal{A}_{(k,s_i)} - \mathcal{C}_{(k,s_i)}^{\top} \mathcal{B}_{(k,s_i)}^{-1} \mathcal{C}_{(k,s_i)},   
    \end{equation}
    \begin{equation}\label{eq:fh-x-t}
        \mathcal{X}_{(T,s_i)} = Q,    
    \end{equation}
    \vspace*{-5mm}
    \begin{align}\label{eq:fh-a}
        \mathcal{A}&_{(k,s_i)} = Q + \sum\nolimits_{h=0}^{L-\xi_k}q_{ih}\sum\nolimits_{r=1}^{h} A^{r \top} Q A^{r} \,+ \\
        & \sum\nolimits_{h=0}^{L-\xi_k} \sum\nolimits_{j=1}^{N} \zeta_{(i,h,j)} (A^{h+1})^{\top} \mathcal{X}_{(k+1+h,s_j)} A^{h+1}, \notag 
    \end{align}
    \vspace*{-6mm}
\begin{align}\label{eq:xik}
\begin{aligned}
    \xi_k \triangleq \max \{0, k+1+L-T \}
\end{aligned}    
\end{align}
\end{subequations}
so that $\xi_{T-1}=L$, 
and $\xi_{k} = 0$ for all $k<T-L$.

The optimal cost is
\begin{equation}\label{eq:fh-cost}
   \!\! J_{T}^{\star}(x_0) = x_{0}^{\top} \!\! \left( \! \sum\nolimits_{i=1}^{N} \! \vartheta_{i} \, \mathcal{X}_{(0,s_i)} \,\!\!\! \right) \! x_{0} \!+\!\!
    \sum\nolimits_{i=1}^{N} \! \vartheta_{i} \, g_{(0,s_i)}, \!\!
\end{equation}
where $\{\vartheta_{i}\}$ indicates the initial probability distribution of the FSMC's states, and, by convention, $\theta_{-1} = \theta_{0}$, so that
\begin{equation}\label{eq:vartheta-i}
    \vartheta_{i} \triangleq \mathbb{E}\left( 1_{\{\theta_{0}=s_{i}\}} \right) = \mathbb{E}\left( 1_{\{\theta_{-1}=s_{i}\}} \right),
\end{equation}
\vspace*{-7mm}
\begin{align}\label{eq:fh-gk}
    & g_{(k,s_i)} = \sum\nolimits_{h=0}^{L-\xi_k} \sum\nolimits_{j=1}^{N} e_{i}^{\top}
    \varsigma_{(h,j)} \bigg( \sum\nolimits_{r=1}^{h} \sum\nolimits_{\nu=0}^{r-1} \notag \\
    & \mathop{\mathrm{tr}}( A^{\nu \top} Q A^{\nu} \Sigma_W ) + 
    \sum\nolimits_{\nu=0}^{h} \mathop{\mathrm{tr}}(A^{\nu \top} \mathcal{X}_{(k+1+h,s_j)} A^{\nu} \Sigma_W ) \,+ \notag \\
    & g_{(k+1+h,s_j)}\bigg), \quad  g_{(T,s_i)} = 0.
\end{align}
\begin{pf}
    See the Appendix.
\end{pf}
\end{thm}

\section{Closed-loop system stability analysis}\label{sec:stability}
An infinite-horizon control strategy aims at guaranteeing the convergence of the system's state to an equilibrium point. 
This goal is achievable only for the stabilizable systems, formally defined as follows.
\begin{defn}[Stabilizability]\label{def:stabiliz} 
A system \eqref{eq:state}--\eqref{eq:p-ij} is stabilizable with one time-step delayed actuation link state observation if, for any initial condition $(x_0,\theta_0)$, and each link state $s_{\ell} \in \mathcal{S}$, there exists a link state-dependent gain $K_{(\infty,s_{\ell})}$, such that  $u_{k}^{c}=K_{(\infty,\theta_{k-1})}x_k$ is the stabilizing SF control input for \eqref{eq:state}.
\end{defn}
\begin{rem}\label{rem:stability}
The FSMC link model yields an MJLS model based on \eqref{eq:xtau}, and for MJLSs, the notions of MS stability, exponential MS stability, and stochastic stability are equivalent \cite{costa2006discrete,zacchialun2019automatica}. Thus, we recall only the definition of the MS stability.
\end{rem}
\begin{defn}[Mean square stability]\label{def:mss}
The system \eqref{eq:state}--\eqref{eq:p-ij} 
is mean-square stable if there exist equilibrium points $x_{e}$ and $X_{e}$, independent from the initial condition $(x_0,\theta_0)$, such that the following holds $\forall (x_0,\theta_0)$:
    \begin{equation}\label{eq:mss}
        \text{\small $\lim_{k \to \infty} \|\mathbb{E} (x_k)-x_{e} \| = 0,~~
        \lim_{k \to \infty} \|\mathbb{E} (x_k x_k^{\top} )-X_{e}\| = 0.$}
    \end{equation}
\end{defn}
In \eqref{eq:mss}, $\|\cdot\|$ indicates an arbitrary matrix norm.

\subsection{MJLS model for stability analysis}\label{subsec:stability-model}
\begin{figure}
\begin{center}
\includegraphics[width=\columnwidth]{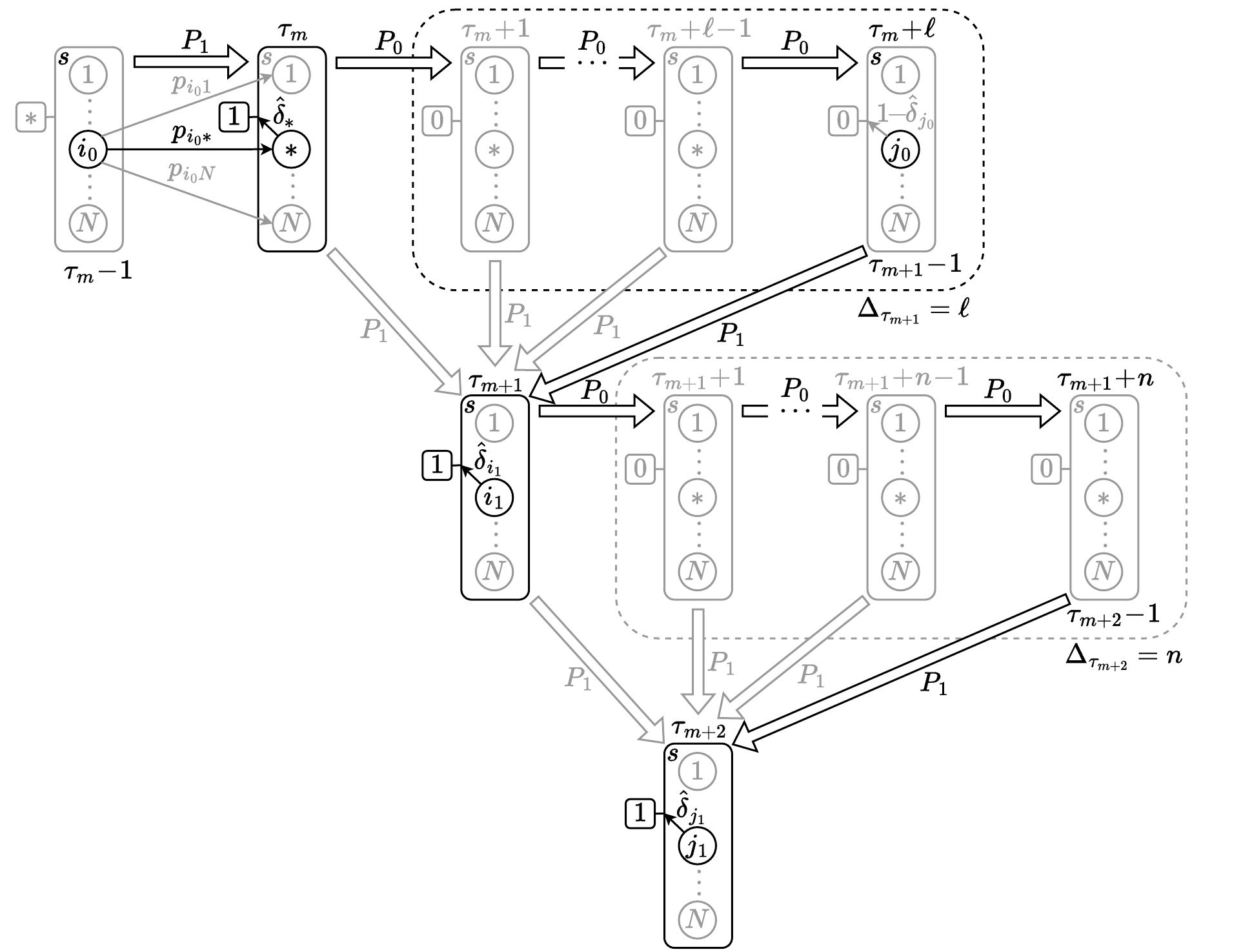}
\caption{The components of the augmented MJLS's discrete states for stability analysis. 
The solid rectangular boxes group the states of the FSMC (represented by circles) in specified time instances; the solid square boxes indicate the transmission outcomes, and the dashed rectangular boxes group the sequences of consecutive message dropouts. The thin arrows pinpoint the probabilities of specific events, while the thick arrows enclose the probabilities of all possible relevant events. The asterisks mark arbitrary values within their admissible sets.}\label{fig:stability}
\end{center}
\end{figure}

The system \eqref{eq:z-k}--\eqref{eq:tp} describes the behavior of the system \eqref{eq:state}--\eqref{eq:p-ij} from the (remote) controller perspective. Stability analysis, however, requires a different perspective, considering each control message transmission outcome as the actuators see it. So, to apply Definition \ref{def:mss}
to the system \eqref{eq:z-k}--\eqref{eq:xtau} from Proposition \ref{prop:equiv}, couch the system in a Markovian framework by considering the augmented state $(x_{\tau_{(m+1)}},\varphi_{\tau_{(m+1)}})$, with 
\begin{subequations}\label{eq:varphi}
\begin{equation}\label{eq:varphi-current}
    \varphi_{\tau_{(m+1)}} \triangleq (\theta_{\tau_{(m+1)}},\Delta_{\tau_{(m+1)}},\theta_{\tau_{(m+1)}-\Delta_{\tau_{(m+1)}}-2}).
\end{equation}
Notice from \eqref{eq:tau} that
\begin{equation}\label{eq:varphi-current-component}
    \theta_{\tau_{(m+1)}-\Delta_{\tau_{(m+1)}}-2} = \theta_{\tau_{(m)}-1}.
\end{equation}
The following augmented state is  $(x_{\tau_{(m+2)}},\varphi_{\tau_{(m+2)}})$, with
\begin{equation}\label{eq:varphi-next}
    \varphi_{\tau_{(m+2)}} \triangleq (\theta_{\tau_{(m+2)}},\Delta_{\tau_{(m+2)}},\theta_{\tau_{(m+2)}-\Delta_{\tau_{(m+2)}}-2}),
\end{equation}
\begin{equation}\label{eq:varphi-next-component}
    \theta_{\tau_{(m+2)}-\Delta_{\tau_{(m+2)}}-2} = \theta_{\tau_{(m+1)}-1}.
\end{equation}
\end{subequations}
Fig.~\ref{fig:stability} emphasizes the components of the augmented system's discrete states in black and the relevant transition probabilities for the time instances $\tau_{(m+1)}$ and $\tau_{(m+2)}$. 

From the Markov property, chain rule of the probability, conditioned version of Bayes' theorem, and \eqref{eq:varphi}, 
\begin{align}\label{eq:tps}
    &\mathbb{P}(\varphi_{\tau_{(m+2)}}\!=(s_{j_1},n,s_{j_0}) \mid \varphi_{\tau_{(m+1)}}\!=(s_{i_1},\ell,s_{i_0})) = \\
    & \frac{e_{i_0}^{\top}P_{1}P_{0}^{\ell}e_{j_0}^{}e_{j_0}^{\top}P_{1}e_{i_1}^{}}{e_{i_0}^{\top}P_{1}P_{0}^{\ell}P_{1}e_{i_1}^{}}e_{i_1}^{\top}P_{0}^{n}P_{1}e_{j_1}^{}
    \triangleq \mu_{((i_{1},\ell,i_{0}),(j_{1},n,j_{0}))}.\notag
\end{align}
Notice from \eqref{eq:tp}, \eqref{eq:L}, and \eqref{eq:tps} that for any $(s_{i_1},\ell,s_{i_0})$,
$\sum\nolimits_{j_{1}=1}^{N}\sum\nolimits_{n\in\mathbb{Z}^{\geq}}\sum\nolimits_{j_{0}=1}^{N}
\mu_{((i_{1},\ell,i_{0}),(j_{1},n,j_{0}))} = 1$,
i.e., the system \eqref{eq:xtau} driven by \eqref{eq:tps} is an MJLS.

\begin{lem}\label{lemma:sys-stability}
The system \eqref{eq:state}--\eqref{eq:p-ij} under an arbitrary actuation link-state-dependent control $u_{k}^{c}=K_{(\infty,\theta_{k-1})}x_k$
is an MJLS described by \eqref{eq:z-k}--\eqref{eq:xtau}, \eqref{eq:varphi}, and \eqref{eq:tps}.
\begin{pf}
    It follows from Proposition \ref{prop:equiv}, \eqref{eq:varphi}, \eqref{eq:tps}, and noticing that both $x_{\tau_{(m+2)}}$ and $\varphi_{\tau_{(m+2)}}$ belong to the actuators' information set in $\tau_{(m+2)}$, which allows us to adopt a posteriori view of the MJLS and compute expectations backward in time.
\end{pf}
\end{lem}

The following section uses this MJLS to derive the easy-to-test condition of the closed-loop system stability for any given infinite-horizon control law. 

\subsection{Closed-loop system stability condition}\label{subsec:stability-check}
We follow the standard MJLS approach tailored here for the problem at hand to derive the necessary and sufficient conditions for the closed-loop system stability. 
We highlight the technical details specific to the considered scenario and omit the presentation of the typical steps thoroughly explained in \cite[Ch.~3]{costa2006discrete}.

Consider an MJLS described by \eqref{eq:z-k}--\eqref{eq:xtau} and \eqref{eq:tps} for an arbitrary  infinite-horizon SF control strategy with FSMC-dependent gain $K_{(\infty,\theta_{\tau_{(m)}-1})}$, where 
$m\in\mathbb{Z}^{\geq}$ and $\tau_{(m)}$ obeys \eqref{eq:tau}.
To obtain the recursive difference equation for the second moment of the system’s state, 
let $1_{\{\varphi_{\tau_{(m+1)}}=(s_{i_1},\ell,s_{i_0})\}}$ be the indicator function specifying the membership (or non-membership) of a given element in the set. For notational convenience, let
\begin{equation}\label{eq:second-moment-component}
   M_{({i_1},\ell,{i_0})}^{\tau_{(m+1)}} \!\triangleq \mathbb{E}\!\left( x_{\tau_{(m+1)}} x_{\tau_{(m+1)}}^{\top} \!1_{\{\varphi_{\tau_{(m+1)}}=(s_{i_1},\ell,s_{i_0})\}} \!\right)\!.\!\!
\end{equation}
$\mathbb{E}\!\left( x_{\tau_{(m+1)}} x_{\tau_{(m+1)}}^{\top} \right) \!=\! \sum\nolimits_{i_{1}=1}^{N}\sum\nolimits_{\ell\in\mathbb{Z}^{\geq}}\sum\nolimits_{i_{0}=1}^{N} \!\mathbb{E}\!\left( M_{({i_1},\ell,{i_0})}^{\tau_{(m+1)}} \right)$.

To write expressions concisely, index the values of the ordered triples $\left(v_{1},v_{2},v_{3}\right)$ representing the operational modes of the closed-loop system, with 
$1\leq v_{1},v_{3}\leq N$ and $0\leq v_{2}\leq L$, using an invertible mapping $f:\mathbb{Z}^{3}\to\mathbb{Z}$,
\begin{equation}\label{eq:mapping}
    f\left(v_{1},v_{2},v_{3}\right)=N^{2}v_{2}+N(v_{3}-1)+v_{1}.
\end{equation}
Then, the augmented system's discrete condition $\mathrm{c}_{\mathrm{i}}$ will be a shorthand for $(s_{i_1},\ell,s_{i_0})$, with $f({i_1},\ell,{i_0})=\mathrm{i}$.

Notice from \eqref{eq:tp} and \eqref{eq:L} that for all $\mathrm{i},\mathrm{j}\in \mathbb{Z}^{>}$, such that $f({i_1},\ell,{i_0}) = \mathrm{i}$, $f({j_1},n,{j_0}) = \mathrm{j}$, and $\mathrm{L}\triangleq (L+1)N^2$,
\begin{equation}\label{eq:mu-compact}
    \sum\nolimits_{\mathrm{j}=1}^{\mathrm{L}} \mu_{(\mathrm{i},\,\mathrm{j})} \geq 1 - \epsilon.
\end{equation}
In the standard MJLS setting having system matrices depending on the present and not subsequent operational mode, the MS stability of the system without process noise is equivalent to the MS stability of the system with Gaussian process noise \cite[Th.~3.33]{costa2006discrete}. 
The following sections also show this is the case for the considered setting described by \eqref{eq:xtau} and \eqref{eq:tps}.

\emph{Noiseless setting:} Let $w_k \!=\! 0$ $\forall k\!\in\!\mathbb{Z}^{\geq}$. 
\eqref{eq:xtau}, \eqref{eq:varphi}--\eqref{eq:mapping} $\Rightarrow$
\begin{align}\label{eq:second-moment-comp}
& M_{({j_1},n,{j_0})}^{\tau_{(m+2)}} = \mathbb{E}\left( x_{\tau_{(m+2)}} x_{\tau_{(m+2)}}^{\top} 1_{\{\varphi_{\tau_{(m+2)}}=(s_{j_1},n,s_{j_0})\}} \right) = \notag \\
& \left(A^{n+1} + \mathit{\Psi}_{(n)}K_{(\infty,s_{j_0})} \right) 
\left( \sum\nolimits_{\mathrm{i}=1}^{\mathrm{L}} M_{\mathrm{i}}^{\tau_{(m+1)}} \mu_{(\mathrm{i},(j_{1},n,j_{0}))} \right) \cdot \notag \\
& \left(A^{n+1} + \mathit{\Psi}_{(n)}K_{(\infty,s_{j_0})} \right)^{\top}
\end{align}
$\forall\,\mathrm{i}\in \mathbb{Z}^{>}$ such that $f({i_1},\ell,{i_0}) = \mathrm{i}$ and $\mathrm{L}\triangleq (L+1)N^2$.

\begin{figure}
\center
\includegraphics[width=\columnwidth]{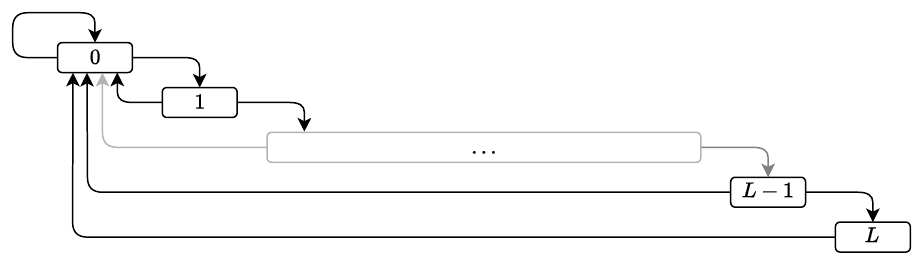}
\caption{Graphical representation of the stochastic process $\{\Delta_{\tau_{(m+1)}}\}$: the rectangles mark the process states labeled with the values each random variable $\Delta_{\tau_{(m+1)}}$ can assume, and the directed arcs between states indicate the non-zero probability transitions from the current to the next state.}\label{fig:delta-state-space}
\end{figure}

For notational convenience, let $\mathcal{M} \triangleq \left[ \mu_{({\mathrm{i},\,\mathrm{j}})}\right]_{\mathrm{i},\,\mathrm{j}=1}^{\mathrm{L}}$ and
\begin{equation}\label{eq:mathcalL}
    \mathcal{L}_{\mathrm{j}} = \mathcal{L}_{({j_1},n,{j_0})} \triangleq A^{n+1} + \mathit{\Psi}_{(n)}K_{(\infty,s_{j_0})}.
\end{equation}
Denote by $\otimes$ the Kronecker product, 
by $\mathop{\mathrm{vec}}$ the matrix vectorization, and
by $\mathop{\mathrm{vec}^{2}}$ the matrix sequence vectorization formally defined as follows: $\forall M_{\mathrm{i}} \in \mathbb{R}^{n_x\times n_x}$,
\begin{equation}\label{eq:vec2}
    \mathop{\mathrm{vec}^{2}}\left( \left( M_{\mathrm{i}} \right)_{\mathrm{i}=1}^{\mathrm{L}} \right) = 
    \text{\small $
    \begin{bmatrix}
        \left(\mathop{\mathrm{vec}}\left(M_{1} \right)\right)^{\top} \cdots \,\left(\mathop{\mathrm{vec}}\left(M_{\mathrm{L}}\right)\right)^{\top}
    \end{bmatrix}^{\top}.
    $}
\end{equation}
Then, the vector form of \eqref{eq:second-moment-comp} is
\begin{equation}\label{eq:second-moment-vec}
    \mathop{\mathrm{vec}^{2}}\text{\small $\left( \!\left( M_{\mathrm{j}}^{\tau_{(m+2)}} \right)_{\mathrm{j}=1}^{\mathrm{L}} \right)$} = \Lambda 
    \mathop{\mathrm{vec}^{2}}\left( \left( M_{\mathrm{i}}^{\tau_{(m+1)}} \right)_{\mathrm{i}=1}^{\mathrm{L}} \right)
\end{equation}
where $\Lambda$
is the MS stability verification matrix, and $I_{n_x^2}$ denotes the identity matrix of size $n_x^2$.
\begin{equation}\label{eq:Lambda}
\!\!
\Lambda \!=\! \text{\small $\left( \!\bigoplus\nolimits_{\mathrm{j}=1}^{\mathrm{L}} \!\!
    \left( \mathcal{L}_{\mathrm{j}} \!\oplus\! \mathcal{L}_{\mathrm{j}} \right)\!\right)\!\!
    \left(\! \mathcal{M}^{\top} \!\!\!\oplus\! I_{n_x^2} \!\right) \!=\! 
    \begin{bmatrix}
        \left( \mathcal{L}_{\mathrm{j}} \!\oplus\! \mathcal{L}_{\mathrm{j}} \right)\!\mu_{(\mathrm{j},\,\mathrm{i})}
    \end{bmatrix}_{\mathrm{i},\,\mathrm{j}=1}^{\mathrm{L}}$}\!
\end{equation}
The following theorem, where $\rho$ indicates the spectral radius of a square matrix, i.e., the largest absolute value of its eigenvalues, provides the origin of the name for the matrix defined by \eqref{eq:Lambda}.
\begin{thm}\label{thm:mss-verif}
    The system \eqref{eq:state}--\eqref{eq:p-ij} with $w_k = 0~\forall k\in\mathbb{Z}^{\geq}$ and any given infinite-horizon control strategy satisfying \eqref{eq:control} is mean-square stable if and only if $\rho\left(\Lambda\right)<1$.    
\begin{pf}
    From Lemma \ref{lemma:sys-stability}, proving the assertion for the system \eqref{eq:z-k}--\eqref{eq:xtau} coupled with \eqref{eq:tps} corresponds to proving it for the system \eqref{eq:state}--\eqref{eq:p-ij} combined with \eqref{eq:control}. 
From \eqref{eq:z-k}--\eqref{eq:xtau} and \eqref{eq:varphi}--\eqref{eq:vec2}, the second moment of the MJLS evolves according to \eqref{eq:second-moment-vec}, where $\Lambda$, as defined by \eqref{eq:Lambda}, is a fixed matrix. Hence, the standard MJLS approach holds and applying the steps from the proof of \cite[Th.~3.9]{costa2006discrete} directly leads to the desired result. $\qed$
\end{pf}
\end{thm}
\emph{Contribution of the Gaussian process noise:}
Let 
\begin{equation}\label{eq:noise-second-moment}
    \mathcal{W}_{\mathrm{j}} = \text{\small $\mathcal{W}_{({j_1},n,{j_0})} \triangleq 
    \sum\nolimits_{h=0}^{n} \left(A^{n-h}\right) \Sigma_W \left(A^{n-h} \right)^{\top}$},
\end{equation}
\vspace*{-5mm}
\begin{equation}\label{eq:mode-probability}
    \psi_{\mathrm{i}}(\tau_{(m+1)}) \triangleq \text{\small $\mathbb{E}\!\left( 1_{\{\varphi_{\tau_{(m+1)}}=\,\mathrm{c}_{\mathrm{i}}\}} \right)$},
\end{equation}
\begin{equation}\label{eq:mathcalG}
    \mathcal{G}_{\mathrm{j}}(\tau_{(m+1)}) \triangleq \text{\small $\left(\sum\nolimits_{\mathrm{i}=1}^{\mathrm{L}} \psi_{\mathrm{i}}(\tau_{(m+1)}) \mu_{(\mathrm{i},\,\mathrm{j})}\right)\mathcal{W}_{\mathrm{j}}$}.
\end{equation}
From \eqref{eq:xtau}, \eqref{eq:varphi}--\eqref{eq:mapping}, \eqref{eq:mathcalL}, and Assumptions \ref{assum:1} and \ref{assum:3},
\begin{equation}\label{eq:second-moment-noise}
M_{\mathrm{j}}^{\tau_{(m+2)}} \!=\! \text{\small $\mathcal{L}_{\mathrm{j}} \!
\left( \sum\nolimits_{\mathrm{i}=1}^{\mathrm{L}} \! M_{\mathrm{i}}^{\tau_{(m+1)}} \! \mu_{(\mathrm{i},\,\mathrm{j})} \!\right)\! 
\mathcal{L}_{\mathrm{j}}^{\top} \!\!+\! \mathcal{G}_{\mathrm{j}}(\tau_{(m+1)})$}.\!\!\!
\end{equation}
Moreover, Assumption \ref{assum:2} combined with \eqref{eq:z-k}--\eqref{eq:tau} and \eqref{eq:varphi} implies the ergodicity of the augmented Markov chain $\left\{\varphi_{\tau_{(m+1)}}\right\}_{m\in\mathbb{Z}^{\geq}}$, as formalized by the following. 
\begin{prop}\label{prop:ergodicity}
If the Markov chain $\left\{\theta_{k}\right\}$ 
is ergodic, then the Markov chain $\left\{\varphi_{\tau_{(m+1)}}\right\}$
is also ergodic.
\begin{pf}
To prove the assertion, show that the DTMC $\{\varphi_{\tau_{(m+1)}}\}$ has a finite number of states and consists entirely of one recurrent and aperiodic class\footnote{See, e.g., \cite{gallager2013stochastic} for the definitions and insights.} if the hypothesis on ergodicity of the Markov chain $\left\{\theta_{k}\right\}$ is satisfied.
From \eqref{eq:varphi}, 
the state space of the DTMC $\{\varphi_{\tau_{(m+1)}}\}$ comprises the states of the finite-state Markov chain $\left\{\theta_{k}\right\}$ observed twice in different suitable time instants defined by \eqref{eq:tau} and the states of the stochastic process $\left\{\Delta_{k}\right\}$ described by \eqref{eq:z-k} and observed in $\tau_{(m+1)}$.
From \eqref{eq:L} and \eqref{eq:mu-compact}, the probability of having $L+1$ consecutive control message dropouts is less than $\epsilon$, the machine epsilon in the  floating-point arithmetic. In other words, the probability of such an event is numerically zero, i.e., negligible in practice. 
Consequently, the stochastic process $\left\{\Delta_{k}\right\}$ has a finite number of states, $L+1$, and may consist entirely of one recurrent and aperiodic class.
Figure \ref{fig:delta-state-space} shows the related graph and allows visually confirming that all the states communicate and are aperiodic if all the depicted arcs are present. 
The transition probabilities between the states of process $\left\{\Delta_{k}\right\}$ depend on the state of the FSMC that evolves according to the Markov chain $\left\{\theta_{k}\right\}$.
Observing that an FSMC cannot have $\hat{\delta}_i$ equal to zero (or one) in all its states $\{ s_i \}_{i=1}^N$, conclude that the ergodic property of $\{\theta_{k}\}$ ensures that all the directed arcs in Figure \ref{fig:delta-state-space} are indeed present, i.e., all the depicted transitions are always possible. A direct consequence is that the finite-state DTMC 
$\{\varphi_{\tau_{(m+1)}}\}$ consists entirely of one class of states that is both recurrent and aperiodic, implying its ergodicity. $\qed$
\end{pf}
\end{prop}
A direct consequence is the following.
\begin{thm}\label{thm:mss-noise}
Under Assumption \ref{assum:2}, the system \eqref{eq:state}--\eqref{eq:p-ij} with any given infinite-horizon control strategy satisfying \eqref{eq:control} is mean-square stable if and only if $\rho\left(\Lambda\right)<1$.
\begin{pf}
By Lemma \ref{lemma:sys-stability}, the system \eqref{eq:z-k}--\eqref{eq:xtau} and \eqref{eq:tps} is trace-equivalent to the system \eqref{eq:state}--\eqref{eq:p-ij} governed by \eqref{eq:control}. 
So, we focus only on the MJLS described by \eqref{eq:z-k}--\eqref{eq:xtau} and \eqref{eq:tps}.
Proposition \ref{prop:ergodicity} implies the existence of the steady-state probability distribution 
$\psi_{\mathrm{i}} \triangleq \lim_{m\to\infty} \psi_{\mathrm{i}}(\tau_{(m+1)})$
independent of the initial distribution $\psi_{\mathrm{i}}(\tau_{(1)})$ so that 
$ \mathcal{G}_{\mathrm{j}} \!\triangleq\! \left(\sum\nolimits_{\mathrm{i}=1}^{\mathrm{L}} \psi_{\mathrm{i}}\,\mu_{(\mathrm{i},\,\mathrm{j})}\right)\!\mathcal{W}_{\mathrm{j}} $. 
Recalling \eqref{eq:second-moment-comp}--\eqref{eq:second-moment-noise} and repeating the steps of the proofs in 
\cite[Sec.~3.4.2]{costa2006discrete} leads to $x_{e}=0$,
    ${ X_{e} = \mathop{\mathrm{vec}^{-2}}\left(\!\left( I_{\mathrm{L}n_x^2} - \Lambda \right)^{-1} 
    \mathop{\mathrm{vec}^{2}}\left( \left( \mathcal{G}_{\mathrm{j}} \right)_{\mathrm{j}=1}^{\mathrm{L}} \right)\!\right) }$
when $\rho\left(\Lambda\right)<1$, and the desired result follows. $\qed$
\end{pf}
\end{thm}

Section \ref{sec:examples} extensively validates Theorem \ref{thm:mss-noise} on a numerical case study, while the following section focuses on designing an optimal infinite-horizon SF control strategy for a generalized dropout compensation.

\section{Optimal infinite-horizon LQR}\label{sec:inf-hor-sol}
If the system \eqref{eq:state} is stabilizable according to Definition \ref{def:stabiliz}, the coupled difference Riccati equations (CDREs)
\eqref{eq:fh-x-a-b-c} converge, resulting in the coupled algebraic Riccati equations (CAREs), with the indices $k$ and $k+1+h$ substituted by $\infty$, and $\xi_{\infty} = 0$ since there is no end of the time horizon. Similarly to the standard MJLS case, at most, one stabilizing solution of CAREs exists, which coincides with the maximal solution of an equivalent convex programming problem. The proof of the uniqueness and asymptotic convergence is on the lines of \cite[Appendix A---Thms A.10 and A.12, Lemma A.14, and Prop. A.23]{costa2006discrete} and does not present technical challenges specific to the wireless networked control scenario. Thus, this section omits the detailed proof of the uniqueness and asymptotic convergence and directly presents the solution to the CAREs in terms of LMIs:
\begin{subequations}\label{eq:lmis}
\begin{equation}\label{eq:lmi-sol}
    \check{\mathcal{X}}_{(s_i)} = \mathop{\mathrm{arg\,max}}_{\mathcal{X}_{(\infty,s_i)}} 
    \mathop{\mathrm{tr}}\left(\sum\nolimits_{i=1}^{N}\mathcal{X}_{(\infty,s_i)}\right) 
\end{equation}
subject to
\begin{align}\label{eq:lmi-core}
    \begin{bmatrix}
        -\mathcal{X}_{(\infty,s_i)} + \mathcal{A}_{(\infty,s_i)} & & \mathcal{C}_{(\infty,s_i)}^{\top} \\
        \mathcal{C}_{(\infty,s_i)} & & \mathcal{B}_{(\infty,s_i)}
    \end{bmatrix} \succeq 0,
\end{align}
\begin{equation}\label{eq:lmi-supp}
    \mathcal{X}_{(\infty,s_i)}\succeq 0, \quad \mathcal{B}_{(\infty,s_i)} \succ 0, 
\end{equation}
\end{subequations}
with the terms in \eqref{eq:lmis} defined by \eqref{eq:fh-x-a-b-c} for $k=\infty$, 
and $\xi_{\infty} = 0$ for all $s_i \in \mathcal{S}$.

For notational conciseness, refer to \eqref{eq:fh-c} and \eqref{eq:fh-b} as $\check{\mathcal{C}}_{(s_i)}$ and $\check{\mathcal{B}}_{(s_i)}$ when their expressions involve the solution, $\left\{\check{\mathcal{X}}_{(s_i)}\right\}$, of the LMIs \eqref{eq:lmis}.

To provide a closed-form expression of the long-run average cost, define the initial probability distribution of the packet error bursts that end in a specific FSMC's state: 
\begin{subequations}\label{eq:pi-n-j}
\begin{equation}\label{eq:pi-n-j-0}
    \pi_{(h,j)}(0) \triangleq \sum\nolimits_{i=1}^{N} \vartheta_{i} e_{i}^{\top} \varsigma_{(h,j)}
\end{equation}
results in $\{\pi_{(h,j)}(0)\}$. It evolves as follows.
\begin{equation}\label{eq:pi-n-j-k}
    \pi_{(\ell,i)}(k+1) \triangleq \sum\nolimits_{h=0}^{L} \sum\nolimits_{j=1}^{N} \pi_{(h,j)}(k) e_{j}^{\top} \varsigma_{(\ell,i)}.
\end{equation}
Notice (from the proof of Proposition \ref{prop:ergodicity}) that Assumption \ref{assum:2} implies the existence of the steady-state distribution: 
\begin{equation}\label{eq:pi-n-j-inf}
    \pi_{(\ell,i)} \triangleq \lim_{k\to\infty}\pi_{(\ell,i)}(k).
\end{equation}
\end{subequations}
\begin{thm}\label{thm:inf-hor-lqr}
Given the solution $\left\{\check{\mathcal{X}}_{(s_i)}\right\}$ of the LMIs \eqref{eq:lmis} under Assumption \ref{assum:2}, the resulting infinite-horizon LQR law defining \eqref{eq:control-inf} for a stabilizable system \eqref{eq:state} is
\begin{subequations}
\begin{equation}\label{eq:lmi-u}
    \check{u}_{k}^{c} = K_{(\infty,\theta_{k-1})} x_k,
\end{equation}
\begin{equation}\label{eq:lmi-k}
    K_{(\infty,s_i)} = - \check{\mathcal{B}}_{(s_i)}^{-1}\check{\mathcal{C}}_{(s_i)}
\end{equation}
for $\theta_{k-1}=s_i$, and the optimal cost that minimizes \eqref{eq:cost-def-inf}
\begin{align}\label{eq:lmi-cost}
   J&_{\!\infty}^{\star} \!=\!
    \sum\nolimits_{h=0}^{L} \sum\nolimits_{j=1}^{N} \pi_{(h,j)} \bigg( \sum\nolimits_{r=1}^{h} \! \sum\nolimits_{\nu=0}^{r-1} \! \notag \\ 
    & \mathop{\mathrm{tr}}( A^{\nu \top} \! Q A^{\nu} \Sigma_W \,\!\!) + \sum\nolimits_{\nu=0}^{h} \!
    \mathop{\mathrm{tr}}(A^{\nu \top} \!\check{\mathcal{X}}_{(s_j)} A^{\nu} \Sigma_W \,\!\!) \bigg).\!
\end{align}
\end{subequations}
\begin{pf}
   This proof is similar to the proof of \cite[Th. 4.6]{costa2006discrete}.
The LQR law in \eqref{eq:lmi-u} complies with \eqref{eq:control} so that $\check{u}_{k}^{c}\in \mathcal{U}_{\infty}$.
From the general MJLSs theory outlined at the beginning of Section \ref{sec:inf-hor-sol}, the stabilizability of the system \eqref{eq:state} ensures that the CDREs \eqref{eq:fh-x-a-b-c} converge into the CAREs. If the LMIs' \eqref{eq:lmis} solution exists, it provides the maximal solution to the CAREs, which coincides with the unique stabilizing solution. Then, \eqref{eq:lmi-k} follows. 
From \eqref{eq:cost-def-inf}, \eqref{eq:zeta-long}, \eqref{eq:tp}, \eqref{eq:fh-cost}, \eqref{eq:fh-gk}, \eqref{eq:pi-n-j}, and $\{\mathcal{X}_{(k,s_i)}\} = \left\{\check{\mathcal{X}}_{(s_i)}\right\}$ for all $k$,
\begin{align*}\label{eq:}
    { J } & { _{\!\infty}^{\star} \!=\! \limsup_{T\to\infty} \frac{1}{T}\!\sum\nolimits_{k=0}^{T-1} \! \sum\nolimits_{h=0}^{L}\!\sum\nolimits_{j=1}^{N} \!\pi_{(h,j)}(k) \bigg( \!\sum\nolimits_{r=1}^{h} }  \notag \\ 
    & { \! \sum\nolimits_{\nu=0}^{r-1} \mathop{\mathrm{tr}}( A^{\nu \top} \! Q A^{\nu} \Sigma_W \,\!\!) \!+\!\! \sum\nolimits_{\nu=0}^{h} \!
    \mathop{\mathrm{tr}}(A^{\nu \top} \!\check{\mathcal{X}}_{(s_j)} A^{\nu} \Sigma_W \,\!\!) \!\bigg),\! }
\end{align*}
which results in \eqref{eq:lmi-cost} and concludes the proof. \qed
\end{pf}
\end{thm}
\begin{rem}
    Both finite- and infinite-horizon 
    gains \eqref{eq:fh-k} and \eqref{eq:lmi-k} depend on the dropout compensation factor $\mathit{\Phi}$. Section \ref{subsec:dropout-factor} examines the impact of the dropout compensation factor on the closed-loop stability and control cost in a numerical case study. An analytical derivation of the optimal $\mathit{\Phi}$ minimizing the control cost with or without stability constraints is an important future research direction enabled by the results of this paper. To this end, the approach of \cite{fu2015optimal} to the optimal dropout compensator design is a good starting point, from which the tedious matrix derivation over structured matrices of numerous terms involving matrix products of structured and unstructured matrices, some of which are elevated to arbitrarily high powers, must be addressed in the FSMC-state-dependent control setting.
\end{rem}

\section{Numerical case study}\label{sec:examples}
This section numerically validates the theoretical results of Theorems \ref{thm:1}, \ref{thm:mss-noise}, and \ref{thm:inf-hor-lqr} on the case study of the rotary inverted pendulum (controlled remotely through a wireless link) chosen to thoroughly examine the impact of a scalar general dropout compensation factor $\phi_1 \in [0,1]$. Notice that $n_u = 1$ implies that $\mathit{\Phi}=\phi_1$, i.e., a scalar. This choice allows us to plot the spectral radius of the MS stability verification matrix and long-run average cost as a function of $\phi_1$ in Figs. \ref{fig:stability-coeff} and \ref{fig:cost-coeff}.

\subsection{System model and parameters}\label{subsec:pendulum}
The pendulum model and parameters are from \cite{rotpen2020}.
The system state consists of the rotary arm and pendulum angles and their derivatives, i.e., the corresponding angular velocities.
The linearization around the unstable equilibrium point and the zero-order hold discretization with a sampling rate of $12$ Hz produces the following discrete-time system matrices.
\begin{equation*}
A = \text{\footnotesize $
\begin{bmatrix}
1 & 0.224 & 0.055 & 0.004 \\
0 & 1.369 & -0.028 & 0.090 \\
0 & 4.994 & 0.391 & 0.167 \\
0 & 8.618 & -0.634 & 1.270
\end{bmatrix}$},~
B = \text{\footnotesize $
\begin{bmatrix}
0.227 \\
0.218 \\
4.944 \\
4.820 
\end{bmatrix}
$}.
\end{equation*}
This linear model is suitable for the ISA100.11a communication protocol \cite[Clause 9.1.9.1.3]{IEC62734:2014} and holds for the small angles from the vertical, e.g., less than $0.175$ rad. 

Consider the system affected by a Gaussian white process noise with a covariance matrix $\Sigma_w \!=\! 2.5\cdot 10^{-9} \, I_{4}$. 
The state-weighting and input-weighting matrices defining the LQR costs are $Q = \bigoplus\{1,5,1,1\}$ and $R = 10$.
The controller aims to balance the pendulum in the upright position corresponding to the inverted pendulum angle equal to zero at the lowest cost.

The controller sends the messages to actuators via a wireless link modeled as the following FSMC.
\begin{equation}\label{eq:tpm2}
P_c = \text{\footnotesize $
\begin{bmatrix}
0.257 & 0.027 & 0.032 & 0.684 \\
0.182 & 0.023 & 0.028 & 0.767 \\
0.172 & 0.022 & 0.027 & 0.779 \\
0.058 & 0.010 & 0.012 & 0.920
\end{bmatrix}$},~
\hat{\delta} = \text{\footnotesize $
\begin{bmatrix}
0.026 \\ 0.375 \\ 0.634 \\ 0.995
\end{bmatrix}^{\top}$}.
\end{equation}
This FSMC accurately describes the successful control message deliveries over IEEE 802.15.4-based links under Wi-Fi interference \cite{10531715} and derives from a wireless networked control system co-design framework 
in \cite{zacchialun2020infocom}.  

\subsection{Finite-horizon LQR example}\label{subsec:fin-hor-case}
\begin{figure}
\begin{center}
\includegraphics[width=\columnwidth]{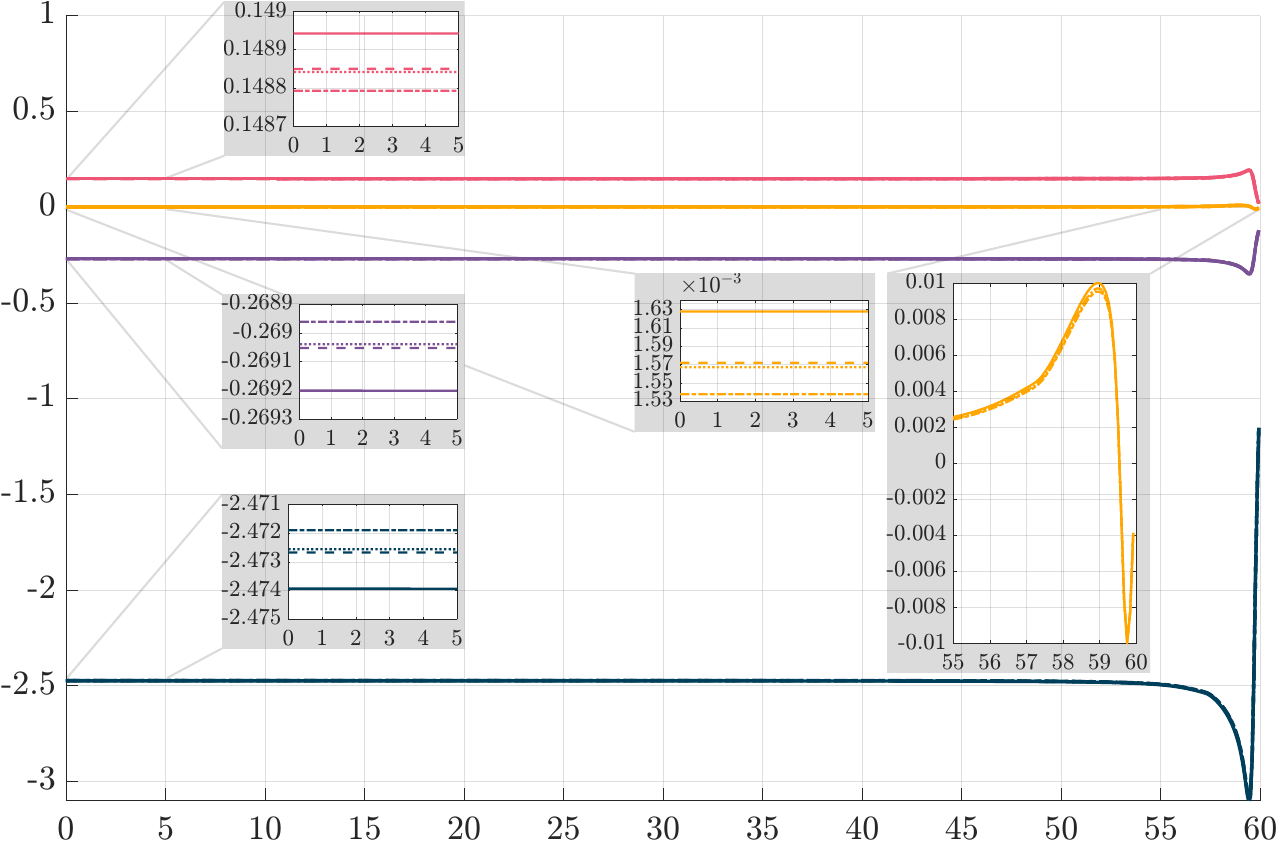}
\caption{Values of the finite-horizon SF gain components in time: solid lines indicate the gain components for the fourth FSMC state, $s_4$, dashed lines refer to the third state, $s_3$, dotted lines identify the components of gain for the second channel state, $s_2$, and dash-dotted lines represent the gain components for the first FSMC state, $s_1$. Orange color identifies the gain component acting on the rotary arm angle, blue singles out the component affecting the pendulum angle, violet indicates the component adjusting the rotary arm angular velocity, and red pinpoints the component varying the pendulum angular velocity. Viewing the plot backwards shows a transitory of around five seconds followed by a steady convergence in the remaining time.}\label{fig:fin-hor-gains}
\end{center}
\end{figure}

To showcase the finite-horizon control strategy in Theorem \ref{thm:1}, consider the time horizon $T = 60$ seconds (corresponding to $720$ samples) and the dropout compensation factor $\phi_1 = 0.921$. See Section \ref{subsec:dropout-factor} for a motivation for using this value of $\phi_1$ as the one producing the most stable behavior in infinite-horizon control setting.
From \eqref{eq:fh-k}, the initial SF control gains are the following.
\begin{align*}
    K_{(0,1)} \!= \text{\footnotesize $
    [\, 0.001538 \!\!\quad\!\! -2.471896 \!\quad\! 0.148793 \!\!\quad\!\! -0.268961 \,]$}, \\
    K_{(0,2)} \!= \text{\footnotesize $
    [\, 0.001567 \!\!\quad\!\! -2.472563 \!\quad\! 0.148842 \!\!\quad\!\! -0.269040 \,]$}, \\
    K_{(0,3)} \!= \text{\footnotesize $
    [\, 0.001572 \!\!\quad\!\! -2.472673 \!\quad\! 0.148850 \!\!\quad\!\! -0.269053 \,]$}, \\
    K_{(0,4)} \!= \text{\footnotesize $
    [\, 0.001628 \!\!\quad\!\! -2.473934 \!\quad\! 0.148943 \!\!\quad\!\! -0.269201 \,]$}.
\end{align*}
Fig.~\ref{fig:fin-hor-gains} 
depicts the gain components' evolution throughout the time horizon and highlights their convergence to the initial ones, as shown by constant values on the zoomed-in plots for the first 5 seconds (i.e., final seconds backwards in time). It indicates a possible convergence of CDREs to CAREs and the system's stabilizability, as detailed in Section \ref{sec:inf-hor-sol} and confirmed in Section \ref{subsec:inf-hor-case}.

Assuming the uniform initial probability distribution of the FSMC's states (i.e., $\vartheta_i = 0.25$ for all $s_i$) and the initial pendulum angle and angular velocity of around $10^{\circ}$ and $44^{\circ}/s$ so that $x_0 = [0~ 0.174~ 0~ 0.767]^{\top}$, 
the cost of the control given by \eqref{eq:fh-cost} is $J_{T}^{\star}(x_0) = 1988.980076$. The initial system state in the unstable equilibrium point $x_0^{\star} = [0~ 0~ 0~ 0]^{\top}$ would produce $J_{T}^{\star}(x_0^{\star}) = 3.416528$, corresponding to the contribution of the second addend in \eqref{eq:fh-cost} to $J_{T}^{\star}(x_0)$. 

\subsection{Infinite-horizon LQR example}\label{subsec:inf-hor-case}
Notice that the TPM $P_c$ in Section \ref{subsec:pendulum} is fully connected, implying that Assumption \ref{assum:2} is satisfied. 
Consider the dropout compensation factor $\phi_1 = 0.921$ again. 
Solving the LMIs \eqref{eq:lmis} in the Robust Control Toolbox for MATLAB \cite{balas2023robust} via its solver \texttt{mincx} and applying Theorem \ref{thm:inf-hor-lqr} produces the infinite-horizon gains $\{K_{(\infty,s_i)}\}$ equal to the initial SF gains $\{K_{(0,s_i)}\}$ in Section \ref{subsec:fin-hor-case} and the optimal long-run average cost $J_{\infty}^{\star} = 0.007692$. The spectral radius of the MS stability verification matrix \eqref{eq:Lambda} with these infinite-horizon gains is $\rho\left(\Lambda\right) = 0.979943$. By Theorem \ref{thm:mss-noise}, the closed-loop system is mean-square stable.

\subsection{Monte Carlo simulation setup}\label{subsec:monte-carlo}
\begin{figure}
\begin{center}
\includegraphics[width=\columnwidth]{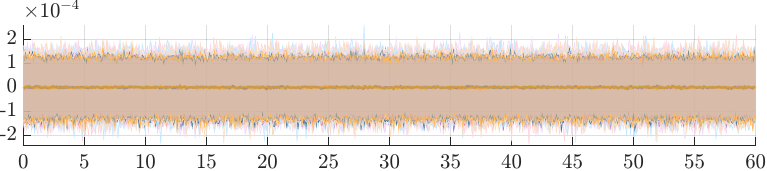}
\caption{Process noise component statistics over $60$ seconds from $500$ independent traces: orange color indicates the component perturbing the rotary arm angle, blue identifies the process noise component affecting the pendulum angle, violet singles out the component disturbing the rotary arm angular velocity, and red pinpoints the component varying the pendulum angular velocity. All median values are close to zero, $95\%$ of observed values are between $\pm 0.0001$, and all the values are greater than $-0.0003$ and less than $0.0003$.}\label{fig:noise-stats}
\end{center}
\end{figure}

\begin{figure}
\begin{center}
\includegraphics[width=\columnwidth]{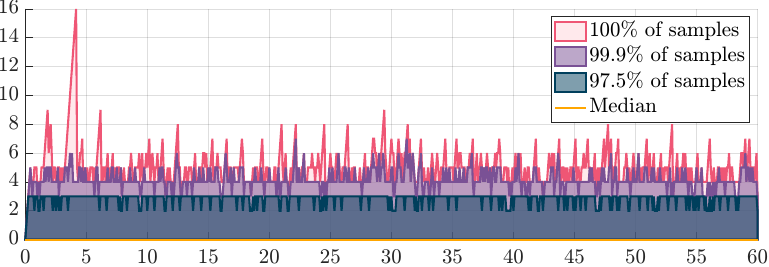}
\caption{Control-packet error burst length statistics over $60$ seconds from a batch of $2000$ traces: the median values (in orange) are all zero, $97.5\%$ of observed values (in blue) are always less than or equal to two, $99.9\%$ (in violet) is less than or equal to $7$, and the maximal values (in red) are predominantly below $8$, but equal $9$ three times, and $16$ once.}\label{fig:link-stats}
\end{center}
\end{figure}

To empirically assess the proposed control strategy, we generated $500$ independent process noise trace samples described in Fig.~\ref{fig:noise-stats} and $2000$ independent control-packet error burst length evolution traces summarized in Fig.~\ref{fig:link-stats}. 
Each packet error burst trace was the output of the FSMC \eqref{eq:tpm2}. Thus, we obtained one million independent samples of the stochastic processes involved by considering every combination of the process noise trace and FSMC evolution. Notably, the packet error burst traces in Fig.~\ref{fig:link-stats} are from one of $1000$ batches of $2000$ traces each, which we selected because of the longest observed burst at the very beginning of the considered time interval of $60$ seconds, a valuable feature for illustrating the system resilience to particularly long consecutive packet error intervals. Furthermore, the selected batch was the only one having $16$ consecutive control packet errors, which was the most extended observed packet error burst\footnote{The most common longest packet error burst in a butch was of length $9$ ($38.6\%$ of batches), closely followed by the one of length $10$ ($35.5\%$); the others were of length $8$ ($7.2\%$), $11$ ($12.8\%$), $12$ ($3.9\%$), $13$ ($1.3\%$), and $14$ ($0.6\%$).}.  

\subsection{Plant dynamics statistics}\label{subsec:pendulum-stat}
To highlight the impact of the control packet dropouts and process noise, we computed the plant's closed-loop dynamics under the proposed infinite-horizon control strategy from Section \ref{subsec:inf-hor-case} starting from the unstable equilibrium point $x_0^{\star} = [0~ 0~ 0~ 0]^{\top}$ as the system's initial state for each stochastic sample from Section \ref{subsec:monte-carlo}. 

\begin{figure}
\begin{center}
\includegraphics[width=\columnwidth]{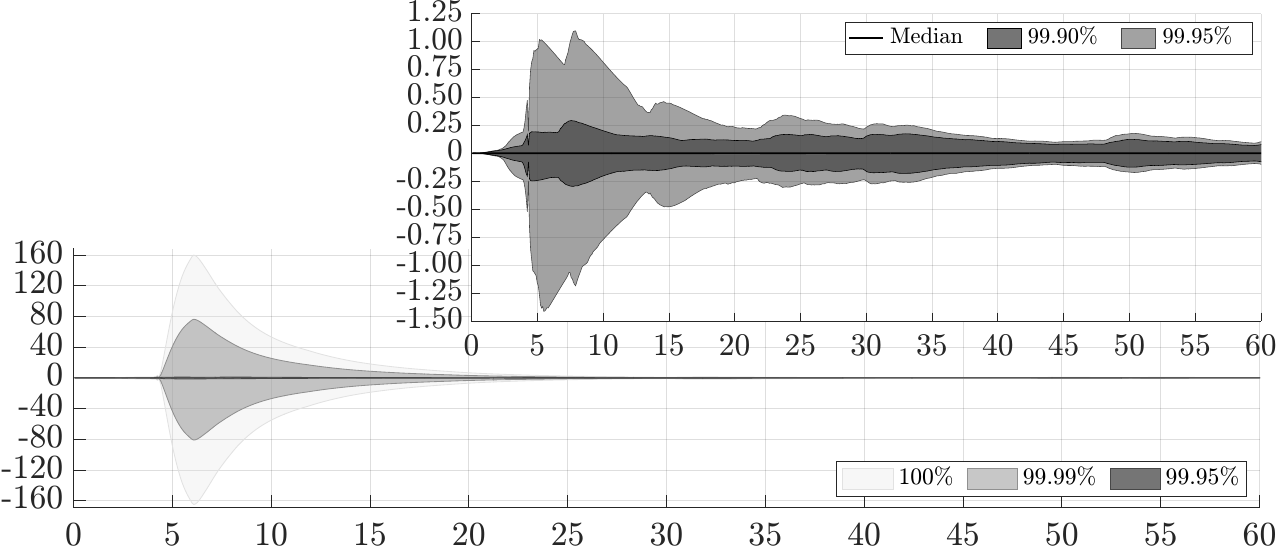}
\caption{Statistics of the rotary arm angle dynamics}\label{fig:dynamics-921-1}
\end{center}
\end{figure}

\begin{figure}
\begin{center}
\includegraphics[width=\columnwidth]{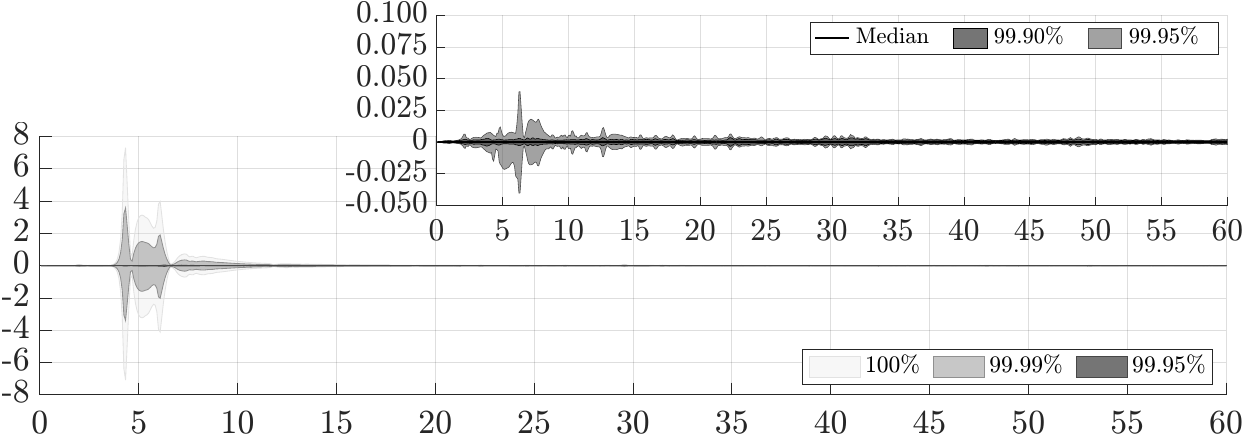}
\caption{Statistics of the pendulum angle dynamics}\label{fig:dynamics-921-2}
\end{center}
\end{figure}

\begin{figure}
\begin{center}
\includegraphics[width=\columnwidth]{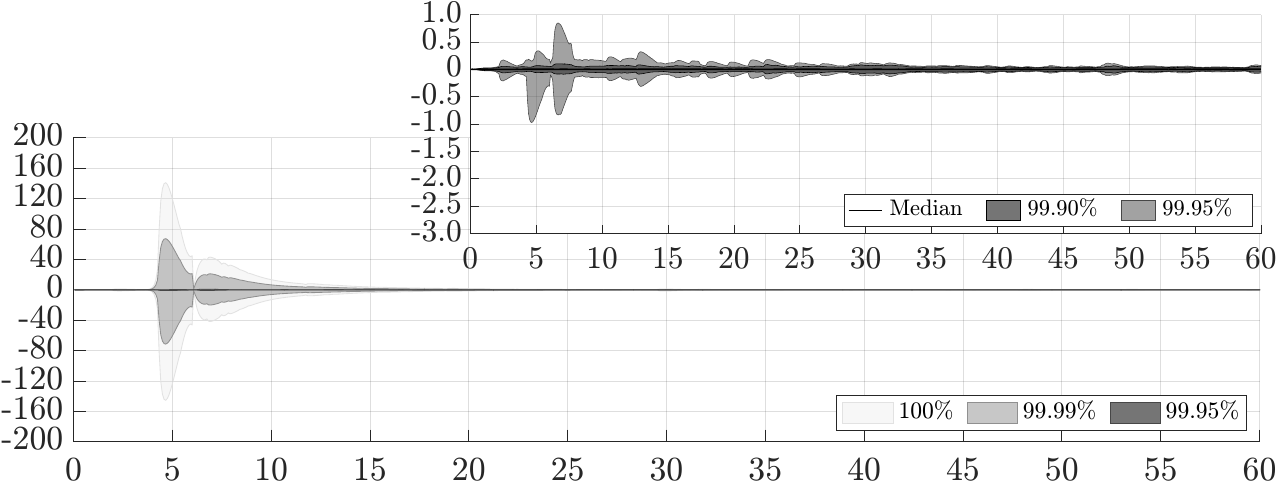}
\caption{Statistics of the rotary arm angular velocity}\label{fig:dynamics-921-3}
\end{center}
\end{figure}

\begin{figure}
\begin{center}
\includegraphics[width=\columnwidth]{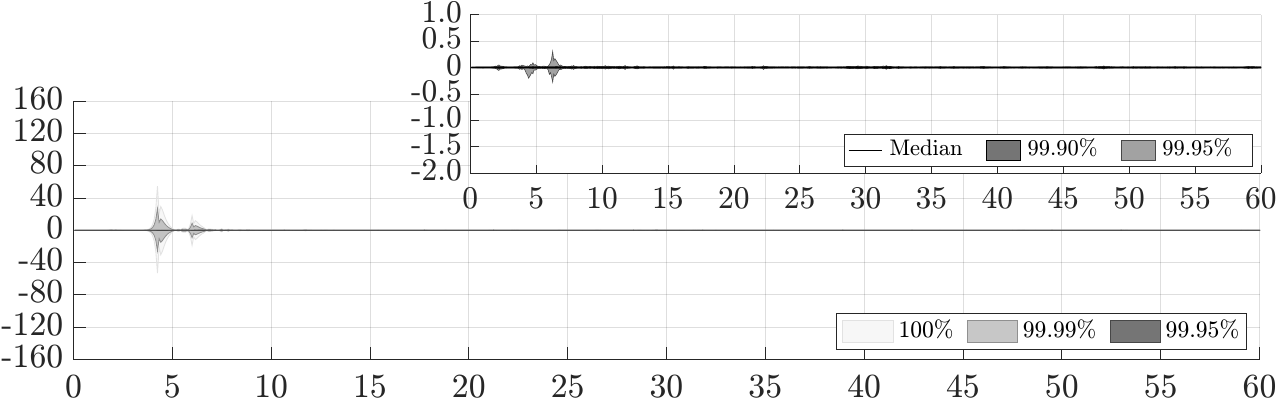}
\caption{Statistics of the pendulum angular velocity dynamics}\label{fig:dynamics-921-4}
\end{center}
\end{figure}

Figs. \ref{fig:dynamics-921-1}--\ref{fig:dynamics-921-4} show the observed system's behavior statistics. The lower part of each figure underlines the effect of the most prolonged packet error burst of length $16$ combined with different realizations of process noise, accounting for $500$ ($0.05\%$ of total) samples. It shows that even under the most unfavorable combination of the control packet dropout and process noise, the closed-loop system dynamics always return to the unstable equilibrium point, as expected from a mean-square stable system. The lower part of Fig. \ref{fig:dynamics-921-2} also indicates that in practical applications, particularly long control packet error bursts may bring the system state outside the valid linearization region ($\pm 0.175$ radians for the rotary inverted pendulum in the exam), an issue outside the MS stability. The upper part of each figure zooms in on the remaining $99.95\%$ of samples, showing that even the improbably long control packet error burst combined with a slight process noise does not create any stability issues, and shorter packet error bursts (combined with any realization of the process noise) do not create any issue either. In particular, under the proposed control strategy, $99.95\%$ of examined pendulum angle traces remain within $0.05$ radians from the vertical, as shown in the upper part of Fig. \ref{fig:dynamics-921-2}.


To assess the long-run average cost, we first exclude as outliers the execution traces obtained with the control packet error burst of length $16$ since the estimated probability of the related event is below $0.0000005$, much lower than the probability of any other considered event. Then, as expected, the $60$-second-run average cost over $99.95\%$ of samples is $0.000163134$, significantly lower than $J_{\infty}^{\star}$ in Section \ref{subsec:inf-hor-case} since it neglects the costly contribution of control packet error bursts of length greater than $9$. Considering also the $250$ less expensive traces with the packet error burst of length $16$ brings the average cost over $99.975\%$ of samples to $0.007333882$, slightly lower than $J_{\infty}^{\star}=0.007691683$. Finally, the $60$-second-run average cost over all samples in the presented butch is $0.110421117$, much higher than the long-run average.

This data confirms that the presented butch represents well the system behavior statistics, identifying both very extreme and typical execution traces. It also corroborates the MS stability of the closed-loop system, thus validating the result of Theorem \ref{thm:mss-noise}.

\subsection{Impact of the dropout compensation factor}\label{subsec:dropout-factor}
\begin{figure}
\begin{center}
\includegraphics[width=\columnwidth]{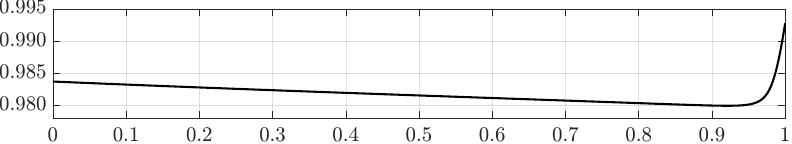}
\caption{The spectral radius of the MS stability verification matrix, $\rho(\Lambda)$, as a function of the dropout compensation factor $\phi_1$ for the rotary inverted pendulum under the proposed infinite-horizon LQR}\label{fig:stability-coeff}
\end{center}
\end{figure}

\begin{figure}
\begin{center}
\includegraphics[width=\columnwidth]{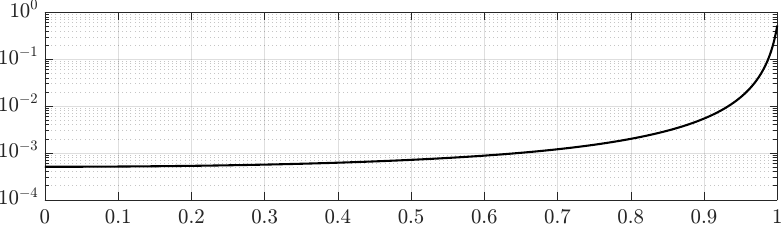}
\caption{Long-run average cost $J_{\infty}^{\star}(\phi_1)$ for the rotary inverted pendulum under the proposed infinite-horizon LQR}\label{fig:cost-coeff}
\end{center}
\end{figure}

An essential application of the presented theoretical results is assessing the closed-loop system stability and control cost for different control strategies and dropout compensation gains.
Fig. \ref{fig:stability-coeff} shows the values of $\rho(\Lambda)$ defined by \eqref{eq:Lambda} 
for varying values of $\mathit{\Phi} = \phi_1$. Notice that $\rho(\Lambda)$ values decrease from $0.983706$ in $0$ to $0.979943$ in $0.921$ and then monotonically increase to $0.992784$ in $1$. This analysis indicates that $\phi_1 = 0.921$ provides the most stable closed-loop behavior in the mean-square sense, ensuring the system's robustness to prolonged control packet error bursts.
Fig. \ref{fig:cost-coeff} presents the long-run average cost of the proposed infinite-horizon LQR for varying values of $\mathit{\Phi} = \phi_1$. This cost 
increases monotonically 
in $\phi_1$, passing from $0.000505$ in $0$ to $0.000511$ in $0.1$, $0.000562$ in $0.3$, $0.000710$ in $0.5$, $0.001195$ in $0.7$, $0.005389$ in $0.9$, and $0.519529$ in $1$.
The long-run average cost and MS stability analyses reveal that the dropout compensation factors between $0$ and $0.921$ provide the trade-off between the two metrics, with particular choices depending on the design's priorities.


The zero-input dropout compensation strategy ($\phi_1 = 0$) provides the lowest long-run average cost for the rotary inverted pendulum. The following section shows that with this dropout compensation, the proposed LQR outperforms the existing wireless control strategies in the MS stability terms (at the price of higher complexity of the CAREs). 

\subsection{Comparative analysis}\label{subsec:comparison}
For the feedback control over lossy communication links with zero-input dropout compensation, \cite{schenato2007proc} and \cite{impicciatore2024tac} presented alternative LQR strategies assuming the packet dropout dynamics are realizations of the Bernoulli or FSMC processes. 

The channel-state-independent LQR gain from \cite{schenato2007proc} relies on the solution of the modified Riccati equation for a specific value of the control packet arrival probability. The successful packet arrival probability for the FSMC \eqref{eq:tpm2} is $0.908862$, producing the following LQR gain for the rotary inverted pendulum from Section \ref{subsec:pendulum}.
\begin{equation*}
    K^B \!= \text{\small $[\, 0.048425 \!\!\quad\!\! -4.419243 \!\quad\! 0.283052 \!\!\quad\!\! -0.495964 \,]$}. \\
\end{equation*}
By \eqref{eq:Lambda}, applying $K^B$ produces $\rho\!\left(\Lambda^{B}\right) = 1.042846$, i.e., unstable system behavior.

The wireless channel-state-dependent LQR gains from \cite{impicciatore2024tac} come from the solution of the CAREs that account only for the expected immediate outcome of the control message transmission without taking explicit care of possible packet error bursts. The resulting gains 
are
\begin{align*}
    K_{(1)}^{M} \!= \text{\small $[\, -0.000285 \!\!\quad\!\! -4.581419 \!\quad\! 0.271732 \!\!\quad\!\! -0.493482 \,]$}, \\
    K_{(2)}^{M} \!= \text{\small $[\, -0.000649 \!\!\quad\!\! -4.446127 \!\quad\! 0.266240 \!\!\quad\!\! -0.483612 \,]$}, \\
    K_{(3)}^{M} \!= \text{\small $[\, -0.000597 \!\!\quad\!\! -4.425511 \!\quad\! 0.265311 \!\!\quad\!\! -0.482016 \,]$}, \\ 
    K_{(4)}^{M} \!= \text{\small $[\, 0.000162 \!\!\quad\!\! -4.229347 \!\quad\! 0.255946 \!\!\quad\!\! -0.466310 \,]$}.
\end{align*}
From \eqref{eq:Lambda}, these gains maintain an MS stability with $\rho\!\left(\Lambda^{M}\right) = 0.999749$.

Finally, the proposed LQR gains from Theorem \ref{thm:inf-hor-lqr} are
\begin{align*}
    K_{(1)}^{P} \!= \text{\small $[\, 0.011000 \!\quad\! -4.528464 \!\quad\! 0.275168 \!\quad\! -0.494649 \,]$}, \\
    K_{(2)}^{P} \!= \text{\small $[\, 0.011208 \!\quad\! -4.528800 \!\quad\! 0.275254 \!\quad\! -0.494734 \,]$}, \\
    K_{(3)}^{P} \!= \text{\small $[\, 0.011242 \!\quad\! -4.528856 \!\quad\! 0.275268 \!\quad\! -0.494748 \,]$}, \\
    K_{(4)}^{P} \!= \text{\small $[\, 0.011634 \!\quad\! -4.529491 \!\quad\! 0.275429 \!\quad\! -0.494908 \,]$}\;
\end{align*}
for $\phi_1 = 0$. Their $\rho\!\left(\Lambda^{P}\right) = 0.983706$, indicating better mean-square stable performance, with faster return to unstable equilibrium point after perturbations from control packet error bursts and process noise. Moreover, the long-run average cost is $0.000505$. 

We numerically validate this analysis via Monte Carlo simulations in Section \ref{subsec:monte-carlo}. 

\begin{figure}
\begin{center}
\includegraphics[width=\columnwidth]{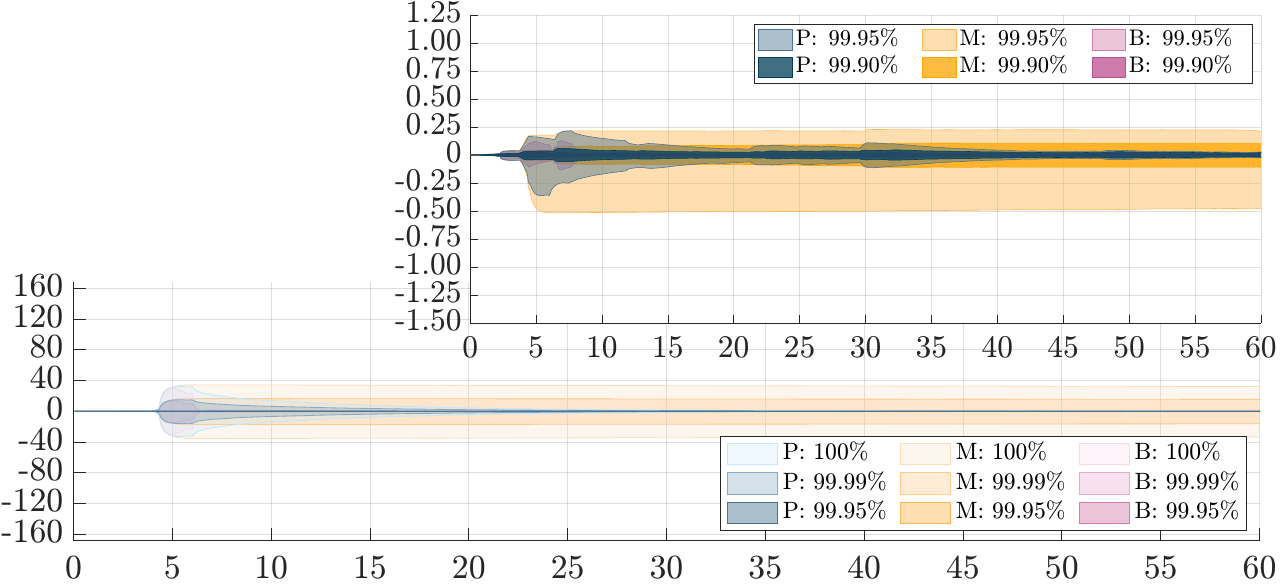}
\caption{Statistics of the rotary arm angle dynamics}\label{fig:dynamics-0-1}
\end{center}
\end{figure}
\begin{figure}
\begin{center}
\includegraphics[width=\columnwidth]{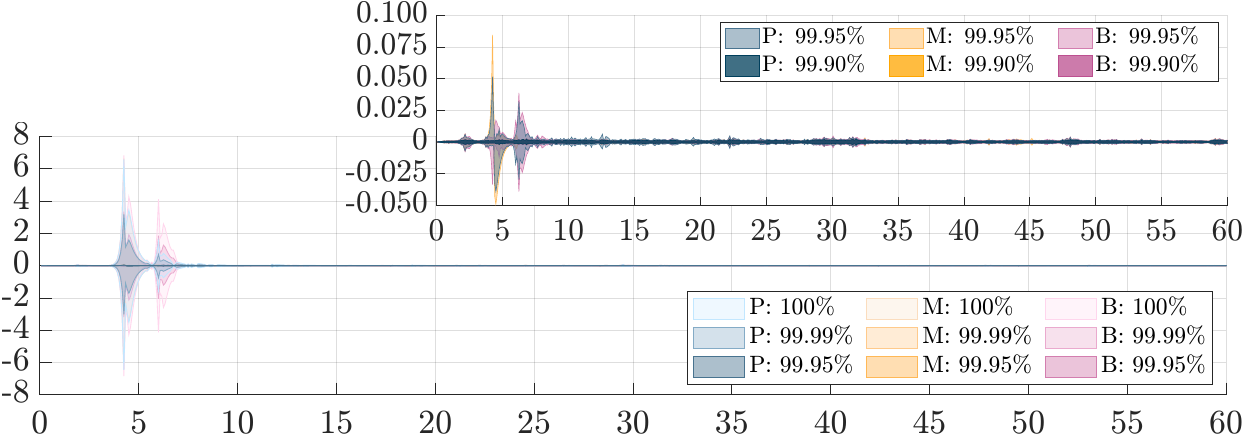}
\caption{Statistics of the pendulum angle dynamics}\label{fig:dynamics-0-2}
\end{center}
\end{figure}
\begin{figure}
\begin{center}
\includegraphics[width=\columnwidth]{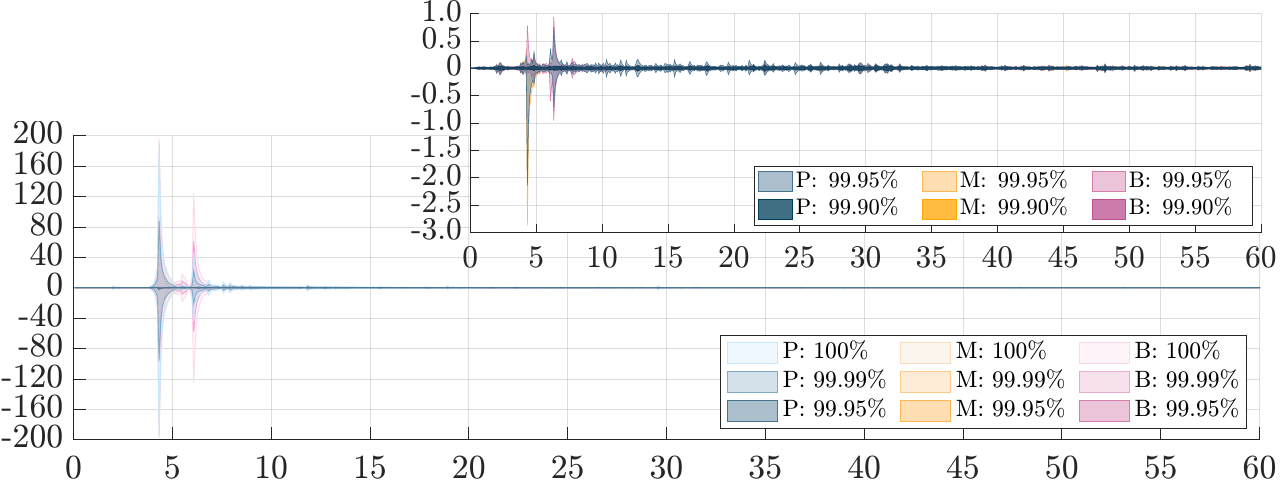}
\caption{Statistics of the rotary arm angular velocity}\label{fig:dynamics-0-3}
\end{center}
\end{figure}
\begin{figure}
\begin{center}
\includegraphics[width=\columnwidth]{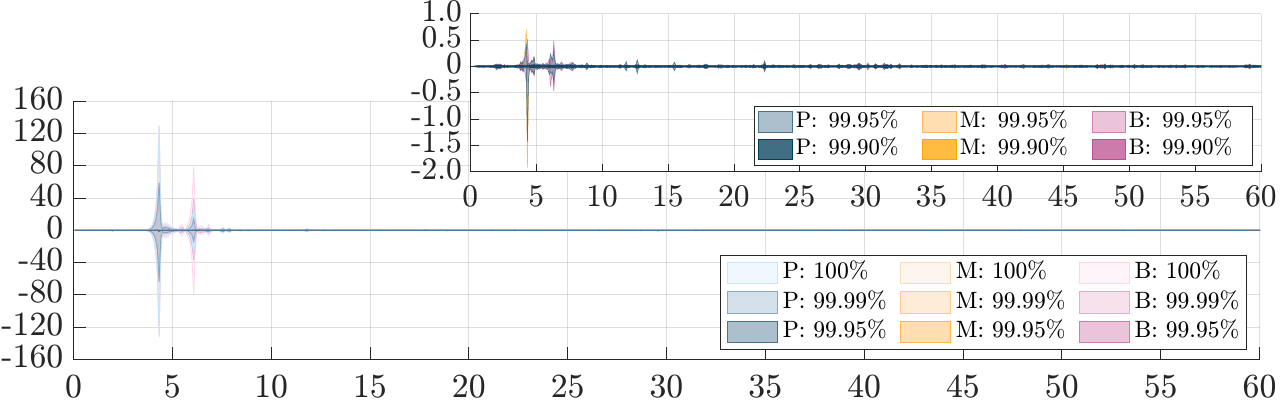}
\caption{Statistics of the pendulum angular velocity dynamics}\label{fig:dynamics-0-4}
\end{center}
\end{figure}

Figs. \ref{fig:dynamics-0-1}--\ref{fig:dynamics-0-4} show the system's behavior statistics for the three LQR strategies, where P indicates the proposed controller (in shades of blue), M identifies the Markovian controller from \cite{impicciatore2024tac} (in shades of orange), and B marks the Bernoullian controller from \cite{schenato2007proc} (in shades of mulberry). Fig. \ref{fig:dynamics-0-1} highlights much faster convergence of the rotary arm angle to the origin for the proposed controller compared to the Markovian. The Bernoullian control strategy exhibits the fastest convergence of the rotary arm angle, which translates into the highest angular velocities and additional oscillations displayed in Fig. \ref{fig:dynamics-0-3}. Moreover, Figs. \ref{fig:dynamics-0-2} and \ref{fig:dynamics-0-4} illustrate that the Bernoullian controller strays the pendulum the most from the unstable equilibrium point and introduces deeper oscillations of the pendulum's angle and velocity, indicating less stable behavior. The proposed controller, instead, maintains the pendulum angle and velocity closer to the origin and efficiently dampens the oscillations.

Similarly to Section \ref{subsec:pendulum-stat}, we computed the $60$-second-run average cost of the LQR strategies over different percentages of closed-loop system evolution traces, summarized in Table \ref{tab:1}. The Bernoullian strategy achieves the lowest average cost over $99.95\%$ of samples that exclude the packet error burst of length $16$. The proposed LQR strategy maintains the best mean-square stable behavior at a lower cost than the average cost of the Markovian strategy. Furthermore, the proposed approach provides the lowest cost over $99.975\%$ and $100\%$ of samples, i.e., those considering packet error bursts longer than $9$.

\begin{table}
    \centering
    \begin{tabular}{|l|c|c|c|}
        \hline
        Controller  \! & $99.95\%$      & $99.975\%$     & $100\%$ \\ \hline
        Bernoullian \!\!\! & $0.000012134$  & $0.001012662$ &  $0.014424479$ \\
        Markovian   \! & $0.000057042$  & $0.005220157$ & $0.078928154$ \\
        Proposed    \! & $0.000015449$  & $0.000892386$ & $0.013577407$ \\ \hline
    \end{tabular}\vspace{1mm}
    \caption{Sixty-second-run average costs of different LQR strategies}\label{tab:1}
\end{table}

Finally, as expected from the analysis in Section \ref{subsec:dropout-factor}, the proposed LQR strategy with the zero-input dropout compensation strategy has a smaller long-run average and observed $60$-second-run average costs compared to the proposed strategy with 
the dropout compensation factor of $0.921$ but presents less stable mean-square behavior, confirmed by the execution traces in Figs. 
\ref{fig:dynamics-921-1}--\ref{fig:dynamics-921-4} and \ref{fig:dynamics-0-1}--\ref{fig:dynamics-0-4} and a higher value of the MS stability verification matrix spectral radius, $\rho(\Lambda)$.

\section{Conclusions}\label{sec:conclusions}
This paper introduced a functional Markov jump linear system modeling of wireless networked control systems with a generalized control message dropout compensation over lossy actuation links modeled by finite-state Markov channels. We envisage extending it to the robust control setting by considering the polytopic time-inhomogeneous Markov channels and different control strategies.

\begin{ack} 
This work was supported in part by the Italian government through the Interministerial Committee for Economic Planning (CIPE) under Resolution 70/2017 (Centre EX-Emerge) and in part by the EU through DigInTraCE under grant 101091801, Resilient Trust under grant 101112282, and the Italian National Recovery and Resilience Plan of NextGenerationEU under project MoVeOver/SCHEDULE with CUP J33C22002880001.
\end{ack}

\bibliographystyle{plain}        
\bibliography{wncs-short}        

\appendix
\section{Appendix: Proof of Theorem \ref{thm:1}} 
\small 
This proof follows the \textit{dynamic programming} approach in Bellman's optimization formulation\footnote{See D.\,P. Bertsekas. \emph{Dynamic programming and optimal control},
volume I \& II. Athena Scientific, Belmont, MA, 1995.}.

Define the cost-to-go for all 
$k\in\mathbb{Z}^{\geq}$ as $J_{T}^{\star}(x_k,\theta_{k-1})$ equal to 
\begin{equation}\label{eq:cost-to-go}
\!\!\!\min_{(u_t^c \in \mathcal{U}_T )_{t=k}^{T-1}}\!\mathbb{E}\bigg(\!\sum\nolimits_{t=k}^{T-1}\!\!\left(x_t^{\top} \!Q x_t \!+\! u_t^{\top}\!Ru_t\right)\! \!+\! x_T^{\top} Q x_T \mid \mathcal{I}_k \!\bigg)\!
\end{equation}
so that $k=0$ provides the optimal cost: from the tower property of the conditional expectation, \eqref{eq:info-set}, \eqref{eq:cost-def-fin},  \eqref{eq:vartheta-i},
\begin{equation}\label{eq:final-cost}
    J_{T}^{\star}(x_0) = \sum\nolimits_{i=1}^{N} \!\vartheta_i \, J_{T}^{\star}(x_0,s_i).
\end{equation}
Consequently, showing that 
\begin{equation}\label{eq:cost-riccati}
    J_{T}^{\star}(x_k,\theta_{k-1}) = x_{k}^{\top}  \mathcal{X}_{(k,\theta_{k-1})} x_{k} + g_{(k,\theta_{k-1})},
\end{equation}
with $\mathcal{X}_{(k,\theta_{k-1})}\succeq 0$ (positive semi-definite) and $g_{(k,\theta_{k-1})}\geq 0$,
is the core of this proof. The main technical challenge lies in transition probabilities \eqref{eq:zeta} defined for the time instances $\tau_{(m)}$ and $\tau_{(m+1)}$ in $\mathcal{T}$, which are unknown to the controller beforehand. 

To obtain the explicit expressions of the cost-to-go and the related optimal SF policy, proceed by \textit{backward induction}. 
The optimal control policy $\check{\mathbf{u}}_T^c$ from \eqref{eq:control-fin} produces the following \textit{terminal cost}. From \eqref{eq:info-set},
\begin{equation}\label{eq:terminal-cost}
    J_{T}^{\star}(x_T,\theta_{T-1}) = \mathbb{E}\left(x_T^{\top} Q x_T \mid \mathcal{I}_T \right) = x_T^{\top} Q x_T.
\end{equation}
Thus, \eqref{eq:cost-riccati} holds in the base case. Moreover, with $\theta_{T-1}=s_i$, \eqref{eq:terminal-cost} implies \eqref{eq:fh-x-t} and the second expression in \eqref{eq:fh-gk} $\forall s_i \!\in\! \mathcal{S}$.

For the \textit{induction step}, assume \eqref{eq:cost-riccati} holds for 
\begin{equation}\label{eq:kp1}
    k+1 = \min \{\tau_{(m+1)}, T\},
\end{equation}
with $\tau_{(m+1)}$ being the time instance of the first successful control message transmission following an arbitrary $k$ 
that may exceed the time horizon $T$. 
Let $\tau_{(m)}$ indicate the time instance of the last successful control message transmission preceding $k+1$. From \eqref{eq:tau} and \eqref{eq:kp1},
\begin{equation}\label{eq:k}
    k = \min\{\tau_{(m)} + \mathit{\Delta}_{\tau_{(m+1)}},\, T-1 \},
\end{equation}
where $\mathit{\Delta}_{\tau_{(m+1)}}$ is a discrete stochastic variable. 
From the tower property of the conditional expectation and \eqref{eq:cost-to-go}, 
\begin{align}\label{eq:cost-taum}
\begin{aligned}
    & J_{T}^{\star}(x_{\tau_{(m)}},\theta_{\tau_{(m)}-1}) = \min_{( u_t^c \in \mathcal{U}_T )_{t=\tau_{(m)}}^{k}} \mathbb{E}\Big(\sum\nolimits_{t=\tau_{(m)}}^{k}\\
    & \qquad\quad \left(x_t^{\top} Q x_t + u_t^{\top}Ru_t\right) +  J_{T}^{\star}(x_{k+1},\theta_{k}) \mid \mathcal{I}_{\tau_{(m)}} \Big).
\end{aligned}
\end{align}

\begin{figure*}[ht]
\raggedright
\begin{align}\label{eq:cost-tau1}
\begin{aligned}
    & \text{\small $J_{T}^{\star}(x_{\tau_{(m)}},\theta_{\tau_{(m)}-1}=s_i) = \!
    \min_{K_{(\tau_{(m)},s_i)}} \Bigg( x_{\tau_{(m)}}^{\top} \Bigg(\mathbb{E} \Big( \sum\nolimits_{r=1}^{k-\tau_{(m)}} A^{r \top} Q A^{r}  + (A^{k-\tau_{(m)}+1})^{\top} \mathcal{X}_{(k+1,\theta_k)} A^{k-\tau_{(m)}+1}\mid \mathcal{I}_{\tau_{(m)}} \Big) \,+ $}\\
    & \quad \text{\small $K_{(\tau_{(m)},s_i)}^{\top} \bigg( \mathbb{E} \Big( \sum\nolimits_{r=1}^{k-\tau_{(m)}} \mathit{\Psi}_{(r-1)}^{\top} Q \mathit{\Psi}_{(r-1)} + \mathit{\Phi}^{r \top} R \mathit{\Phi}^{r} + \mathit{\Psi}_{(k-\tau_{(m)})}^{\top} \mathcal{X}_{(k+1,\theta_k)} \mathit{\Psi}_{(k-\tau_{(m)})} \mid \mathcal{I}_{\tau_{(m)}} \Big) + R \bigg) K_{(\tau_{(m)},s_i)} \,+ $} \\
    & \quad \text{\small  $Q + \mathbb{E} \Big( (A^{k-\tau_{(m)}+1})^{\top} \mathcal{X}_{(k+1,\theta_k)} \mathit{\Psi}_{(k-\tau_{(m)})} K_{(\tau_{(m)},s_i)} + K_{(\tau_{(m)},s_i)}^{\top} \mathit{\Psi}_{(k-\tau_{(m)})} ^{\top} \mathcal{X}_{(k+1,\theta_k)} A^{k-\tau_{(m)}+1} \,+$} \\
    & \quad \text{\small $ \sum\nolimits_{r=1}^{k-\tau_{(m)}}K_{(\tau_{(m)},s_i)}^{\top} \mathit{\Psi}_{(r-1)}^{\top} Q A^{r} +  A^{r \top} Q \mathit{\Psi}_{(r-1)} K_{(\tau_{(m)},s_i)} 
    \mid \mathcal{I}_{\tau_{(m)}} \Big) \! \Bigg) x_{\tau_{(m)}} \,+$} \\
    & \quad \text{\small $\mathbb{E} \Big( \sum\nolimits_{\nu=0}^{k-\tau_{(m)}} \mathop{\mathrm{tr}} (A^{\nu \top}  \mathcal{X}_{(k+1,\theta_k)} A^{\nu} \Sigma_W) +\sum\nolimits_{r=1}^{k-\tau_{(m)}} \!\sum\nolimits_{\nu=0}^{r-1} \mathop{\mathrm{tr}} (A^{\nu \top} \! Q A^{\nu} \Sigma_W) + g_{(k+1,\theta_k)} \mid \mathcal{I}_{\tau_{(m)}} \! \Big)\!\Bigg)$}      
\end{aligned}
\end{align}
\begin{align}\label{eq:cost-tau2}
\begin{aligned}
    & \text{\small $J_{T}^{\star}(x_{\tau_{(m)}},\theta_{\tau_{(m)}-1}=s_i) = 
    x_{\tau_{(m)}}^{\top}\mathcal{A}_{(\tau_{(m)},s_i)}x_{\tau_{(m)}}^{} \!+ g_{(\tau_{(m)},s_i)} \,+$} \\
    & \quad \text{\small $\min_{K_{(\tau_{(m)},s_i)}} \left(\! x_{\tau_{(m)}}^{\top} \left( 
    \mathcal{C}_{(\tau_{(m)},s_i)}^{\top} K_{(\tau_{(m)},s_i)} + K_{(\tau_{(m)},s_i)}^{\top} \mathcal{B}_{(\tau_{(m)},s_i)} K_{(\tau_{(m)},s_i)}^{} + 
    K_{(\tau_{(m)},s_i)}^{\top}\mathcal{C}_{(\tau_{(m)},s_i)}^{} \right) x_{\tau_{(m)}} \right)$}
\end{aligned}
\end{align}
\begin{align}\label{eq:XP}
\begin{aligned}
    & \text{\small $\mathcal{X}_{(\tau_{(m)},s_i)} = \sum\nolimits_{h=0}^{L-\xi_{\tau_{(m)}}} q_{ih} \sum\nolimits_{r=1}^{h} 
    \left( A^{r} + \mathit{\Psi}_{(r-1)} K_{(\tau_{(m)},s_i)} \right)^{\!\top}Q \left(A^{r} + \mathit{\Psi}_{(r-1)} K_{(\tau_{(m)},s_i)}\right) + Q + $} \\
    & \quad\text{\small $\sum\nolimits_{h=0}^{L-\xi_{\tau_{(m)}}} \sum\nolimits_{j=1}^{N} \zeta_{(i,h,j)}
    \left(A^{h+1} + \mathit{\Psi}_{(h)} K_{(\tau_{(m)},s_i)} \right)^{\!\top} \mathcal{X}_{(\tau_{(m)}+h+1,s_j)} \!
    \left(\!A^{h+1} + \mathit{\Psi}_{(h)} K_{(\tau_{(m)},s_i)}\right)\,+$} \\
    & \quad\text{\small $K_{(\tau_{(m)},s_i)}^{\top}\left(R+\sum\nolimits_{h=0}^{L-\xi_{\tau_{(m)}}}
    q_{ih}\sum\nolimits_{r=1}^{h} \mathit{\Phi}^{r \top} R \mathit{\Phi}^r \right)K_{(\tau_{(m)},s_i)}$}
    \end{aligned}
\end{align}
\end{figure*}

From the inductive hypothesis, linearity of the expectation, \eqref{eq:state}, \eqref{eq:info-set}, \eqref{eq:control}, \eqref{eq:tau}--\eqref{eq:xtau}, 
Assumptions \ref{assum:1} and \ref{assum:3}, the cyclic property of the trace, and $w_k$ being a white Gaussian process with zero mean and covariance matrix $\Sigma_w$, we obtain \eqref{eq:cost-tau1}, shown at the top of the page.

Notice that $\tau_{(m)}\leq T-1$.
From \eqref{eq:k} and \eqref{eq:L}, $k-\tau_{(m)}$ is a bounded discrete stochastic variable: $\min(k-\tau_{(m)}) \!=\! 0$, 
\begin{equation}\label{eq:max-k-tau}
    \max(k-\tau_{(m)}) = \min\{L,\,T-1-\tau_{(m)}\}.
\end{equation}
Thus, we define
\begin{equation}\label{eq:xitau}
    \xi_{\tau_{(m)}} \triangleq \max\{0,\tau_{(m)}+1+L-T\}
\end{equation}
so that $\max(k-\tau_{(m)}) = L-\xi_{\tau_{(m)}}$. 

From \eqref{eq:etatau}--\eqref{eq:tp}, the terms independent of $K_{(\tau_{(m)},s_i)}$ and $x_{\tau_{(m)}}$ in \eqref{eq:cost-tau1} become $g_{(\tau_{(m)},s_i)}$ as in \eqref{eq:fh-gk}
since by Assumptions \ref{assum:1} and \ref{assum:3}, the evolution of the process noise is independent of the system's state, control message transmission outcome, and FSMC's state.
On the contrary, the terms that depend on the system's state $x_{\tau_{(m)}}$ evolve according to \eqref{eq:zeta-long}. Explicitly, always from \eqref{eq:etatau}--\eqref{eq:tp}, the terms independent of $K_{(\tau_{(m)},s_i)}$ in \eqref{eq:cost-tau1} form $\mathcal{A}_{(\tau_{(m)},s_i)}$ as in \eqref{eq:fh-a}. 
Defining the terms
$\mathcal{B}_{(\tau_{(m)},s_i)}$ and $\mathcal{C}_{(\tau_{(m)},s_i)}$ as 
in \eqref{eq:fh-b} and \eqref{eq:fh-c} leads to \eqref{eq:cost-tau2}, shown at the top of the page.

Notice that the inductive hypothesis of $\mathcal{X}_{(k+1,\theta_{k})}\succeq 0$ 
for all the values of $\theta_{k}$, together with transition probabilities in \eqref{eq:tp} being nonnegative, $Q\succeq 0$, and $R\succ 0$, implies that $\mathcal{B}_{(\tau_{(m)},s_i)} \succ 0$ and $\mathcal{A}_{(\tau_{(m)},s_i)}\succeq 0$.

Performing the matrix differentiation to find stationary points of $J_{T}^{\star}(x_{\tau_{(m)}},\theta_{\tau_{(m)}-1}=s_i)$ in $K_{(\tau_{(m)},s_i)}$ yields 
\begin{equation}\label{eq:diff}
 \!\!\!2\!\left(\!\mathcal{B}_{(\tau_{(m)},s_i)}K_{(\tau_{(m)},s_i)} \!+ \mathcal{C}_{(\tau_{(m)},s_i)}\! \right) \! x_{\tau_{(m)}}x_{\tau_{(m)}}^{\top} \! = 0,
\end{equation}
which should hold for all possible values of $x_{\tau_{(m)}}x_{\tau_{(m)}}^{\top}$. So, the multiplier of $x_{\tau_{(m)}}x_{\tau_{(m)}}^{\top}$ in \eqref{eq:diff} should be the matrix of all zeros. 
Consequently, $K_{(\tau_{(m)},s_i)}$ is defined by \eqref{eq:fh-k} with $\tau_{(m)}$ in place of $k$. Since  $J_{T}^{\star}(x_{\tau_{(m)}},\theta_{\tau_{(m)}-1}=s_i)$ is a positive quadratic function, its stationary point corresponds to a local minimum. Having only one stationary point for each gain implies that this point constitutes a global minimum. 
Substitute $K_{(\tau_{(m)},s_i)}$ with its expression \eqref{eq:fh-k} in \eqref{eq:cost-tau2} results in $\mathcal{X}_{(\tau_{(m)},s_i)}$ as in \eqref{eq:fh-x} so that 
\begin{equation*}\label{eq:cost-tau3}
    J_{T}^{\star}(x_{\tau_{(m)}},\theta_{\tau_{(m)}-1}=s_i) =\,\!\! g_{(\tau_{(m)},s_i)} + x_{\tau_{(m)}}^{\top}\!\mathcal{X}_{(\tau_{(m)},s_i)}x_{\tau_{(m)}}^{},
\end{equation*}
which implies \eqref{eq:fh-cost} for $\tau_{(m)}=0$ since $\theta_{-1} \notin \mathcal{I}_0$. 

Moreover, \eqref{eq:XP}, shown at the top of the page, proves that $\mathcal{X}_{(\tau_{(m)},s_i)}\succeq 0$.
At last, recall that as a covariance matrix, $\Sigma_w \succeq 0$, so $g_{(\tau_{(m)},s_i)}\geq 0$ since the trace of the product of two positive semi-definite matrices is nonnegative.

Since at each transmission time, the controller selects the SF gain \eqref{eq:fh-k} under the assumption that it will be successfully received, and $\tau_{(m)}$ formalizes this assumption via \eqref{eq:calT}, this proof provides the expressions in $\tau_{(m)}$. Writing the expressions in $k$ instead of $\tau_{(m)}$ results in \eqref{eq:fh-u}--\eqref{eq:fh-gk} and concludes the proof. \qed

\end{document}